\documentclass[10pt,journal]{IEEEtran}
\ifCLASSINFOpdf
\else
\fi
\usepackage{array,booktabs}
\usepackage{algpseudocode}
\usepackage{algorithm}
\usepackage{amsfonts}
\usepackage{amsmath}
\usepackage{amssymb}
\usepackage{amsthm}
\usepackage{graphicx}
\usepackage[noadjust]{cite}
\usepackage{mathrsfs}
\usepackage{stmaryrd}
\usepackage{siunitx}
\usepackage{multicol}
\usepackage{tikz}
\usepackage{setspace}
\usepackage{pstricks, pst-node, pst-plot, pst-circ}
\usepackage{moredefs}
\usetikzlibrary{shapes}

\newcommand{\revise}[1]{\textcolor{black}{#1}}

\DeclareGraphicsExtensions{.png}
\algdef{SE}[DOWHILE]{Do}{doWhile}{\algorithmicdo}[1]{\algorithmicwhile\ #1}%
\algnewcommand{\Initialize}[1]{%
  \State \textbf{Initialize:}
  \Statex \hspace*{\algorithmicindent}\parbox[t]{.8\linewidth}{\raggedright #1}
}

\providelength{\AxesLineWidth}       
\providelength{\plotwidth}           
\providelength{\LineWidth}           
\providelength{\MarkerSize}          
\newrgbcolor{GridColor}{0.8 0.8 0.8}%
\begin{document}
\title{Bayesian Design of Tandem Networks for \\ Distributed Detection With Multi-bit \\ Sensor Decisions}
\author{Alla~Tarighati,~\IEEEmembership{Student Member,~IEEE,}
	 and~Joakim~Jald{\'e}n,~\IEEEmembership{Senior Member,~IEEE,}
\thanks{A.~Tarighati~and~J.~Jald{\'e}n are with the ACCESS Linnaeus Centre, department of signal processing,
KTH Royal Institute of Technology, Stockholm 100 44, Sweden (e-mail: allat@kth.se; jalden@kth.se).

This work has been supported in part by the ACCESS seed project DeWiNe.}
}
\maketitle

\begin{abstract}
We consider the problem of decentralized hypothesis testing under communication constraints in a topology where several peripheral nodes are arranged in tandem. Each node receives an observation and transmits a message to its successor, and the last node then decides which hypothesis is true. We assume that the observations at different nodes are, conditioned on the true hypothesis, independent and the channel between any two successive nodes is considered error-free but rate-constrained. We propose a cyclic numerical design algorithm for the design of nodes using a person-by-person methodology with the minimum expected error probability as a design criterion, where the number of communicated messages is not necessarily equal to the number of hypotheses. The number of peripheral nodes in the proposed method is in principle arbitrary and the information rate constraints are satisfied by quantizing the input of each node. The performance of the proposed method for different information rate constraints, in a binary hypothesis test, is compared to the optimum rate-one solution due to Swaszek and a method proposed by Cover, and it is shown numerically that increasing the channel rate can significantly enhance the performance of the tandem network. Simulation results for $M$-ary hypothesis tests also show that by increasing the channel rates the performance of the tandem network significantly improves.

\end{abstract}

\IEEEpeerreviewmaketitle

\section{Introduction}\label{sec:intro}
\IEEEPARstart{B}{ecause} of reliability, survivability and reduced communication bandwidth requirements, distributed signal processing systems have received significant attention in the past. In the context of distributed detection, considerable progress was made during the past few decades, see \cite{Tsi93,Varsh96,Vis97} and references therein. Distributed detection also regained new interests in relation to wireless sensor networks (WSN) during the past decade. The application of distributed detection in WSNs emerges mainly in communication architecture and resource management \cite{Cham03,Chamb04,Bahe05,Fab10}. For instance, the problem of optimal sensor decisions for a capacity constrained sensor network was studied in \cite{Cham03}, while finding the optimal sensor decisions under global resource constraints was considered in \cite{Chamb04}. In \cite{Bahe05} WSN arranged in serial was considered where channels between the sensors were subject to flat fading. A comprehensive survey of early works in the application of decentralized hypothesis testing in WSNs can be found in \cite{Veer12,Cham07,Chen06}.

In a distributed, or decentralized, hypothesis testing system, observations are made at spatially separated sensors. If the sensors are able to communicate all their data to a central processor there is no fundamental difference from a centralized hypothesis test where the optimal solution is given by threshold tests on the likelihood ratios computed from the complete set of observations. On the other hand, if there are communication constraints on the channels between the sensors, some preliminary processing of the data need to be carried out at each sensor and a compressed, or quantized, version of the received data is then instead given as the sensor output. According to the network arrangement, the output of each sensor is then sent to either another sensor or to a fusion center (FC), which makes the final decision in favor of one of the hypotheses. In the context of distributed detection each sensor is thus an intelligent unit, and is therefore often referred to as a decision maker (or DM) \cite{Varsh96, Vis97}. The goal of this paper is to introduce a general numerical methodology for the design of the DMs in tandem networks for $M$-ary hypothesis testing.

The optimal design of the DMs in a tandem network was previously studied in \cite{VisTT88,Swa93,TanPK91} under the assumption that the observations at the sensors were conditionally independent. This scenario has also recently been generalized in \cite{PenV12} to the case of conditionally dependent observations. Common to \cite{VisTT88,Swa93,TanPK91,PenV12} are that the channels between the DMs are considered to be rate-constrained but error-free. While \cite{VisTT88,Swa93,TanPK91} considered binary hypothesis testing and binary messages between the DMs, \cite{PenV12} relaxed this assumption and considered general $M$-ary hypothesis testing with $M$-valued messages for $M \geq 2$. We shall herein consider $M$-ary hypothesis testing and conditionally independent sensor observations, but will generally allow for higher communication rates than what is provided by $M$-valued messages.

\begin{figure*}[!t]
\begin{center}
\begin{tikzpicture}[align=center,scale=0.75,>=stealth] 
\node (DM1) at (-6,3) [rectangle,minimum size=1.0cm,draw] {DM \\ 1};
\node (dots1) at (-3,3) {$\cdots$};
\node (DMl) at (0,3) [rectangle,minimum size=1.0cm,draw] {DM \\�$l$};
\node (dots2) at (3,3) {$\cdots$};
\node (DMN) at (6,3) [rectangle,minimum size=1.0cm,draw] {DM \\�$N$};
\node (PH) at (0,5.5) [ellipse,inner sep=3mm,draw] {Phenomenon $H$};
\draw [->] (DM1.east) -- (dots1) node  [near start,below,inner sep=6pt] {$u_1$};
\draw [->] (dots1) -- (DMl) node [near start,below,inner sep=6pt] {$u_{l-1}$};
\draw [->] (DMl) -- (dots2) node [near start,below,inner sep=6pt] {$u_{l}$};
\draw [->] (dots2) -- (DMN) node [near start,below,inner sep=6pt] {$u_{N-1}$};
\draw [->] (DMN.east) --(9,3)   node [near end,below,inner sep=6pt] {$u_{N}$};
\draw [->] (PH) -- (DM1.north) node [near end,left,inner sep=9pt] {$x_1$};
\draw [->] (PH) -- (DMl.north) node [near end,right,inner sep=4pt] {$x_{l}$};
\draw [->] (PH) -- (DMN.north) node [near end,right,inner sep=7pt] {$x_{N}$};
\end{tikzpicture} 
\caption{Decentralized hypothesis testing scheme in a serial network.}
\label{fig:tandem}
\end{center}
\end{figure*}
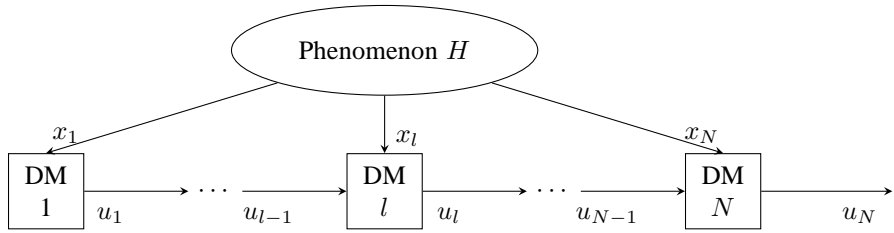

With respect to the optimal performance limits of tandem networks, it was shown in \cite{Tsi93,Papa92} that for distributed networks with two DMs the optimal tandem network performs at least as well as the optimal parallel network. However, when the number of DMs increases parallel networks perform better than serial networks, and for any given distributed detection problem with i.i.d. observations there exists a number of DMs at which the parallel network becomes better \cite{Tsi93}. In the case of a parallel topology with any logical decision functions, the error probability goes to zero very quickly as the number of DMs increases. This does however not hold in general for the tandem topology. It was in fact shown in \cite{TTW08} that the rate of error probability decay of the tandem network is always sub-exponential in the total number of DMs, while the error probability decay of a parallel network is exponential in the total number of DMs \cite{Tsi88}.

The asymptotic performance of parallel and tandem networks has attracted a lot of interest over the past years \cite{Cov69, Tsi88,Papa92, Kop75,TTW08}. It was for instance shown in \cite{Papa92, Cov69} that when the DMs are allowed to send $M$-valued messages for $M$-ary hypothesis testing, a necessary and sufficient condition for the probability of error to asymptotically go to zero is that the log-likelihood ratio of the observation at each DM, between any two arbitrary hypotheses, is unbounded in magnitude. In other words, in the general case with potentially bounded log-likelihood ratios (strictly) more messages than hypotheses are needed to drive the error to zero. In the case of binary hypothesis testing ($M=2$) and for bounded log-likelihood ratios, Cover \cite{Cov69} proposed an algorithm with a four-valued message which achieves zero-limiting probability of error under each hypothesis. This idea was later generalized by Koplowitz \cite{Kop75} to show that $(M+1)$-valued messages are sufficient for achieving zero-limiting probability of error in $M$-ary hypothesis testing, even if the log-likelihood ratios are bounded.

For tandem networks of fixed size, Papastavrou and Athans \cite{Papa92} proposed a simple but suboptimal scheme for the network design in which each DM is optimized for locally minimal error probability at its output, instead of for globally optimal performance. In the particular scheme of \cite{Papa92}, a necessary and sufficient condition to achieve zero-limiting probability of error is also that the log-likelihood ratio of the observation of each DM be unbounded from both above and below. However, a side effect of optimizing the performance (i.e., minimizing the error probability) locally at the output of each DM is that the messages are then again constrained to be $M$-valued for the $M$-ary hypothesis test as a one-to-one relation between the DM output messages and the hypotheses is needed in definition of the local probability of error. Thus, the problem of designing the DMs in a tandem network for arbitrary-valued messages remains largely open \cite{Swa93}, even though it is known that increasing the number of communication messages can improve the performance of a network of sensors arranged in parallel \cite{Lee89}. The latter point was, e.g., exemplified in \cite{IbrH01} where it was shown that allowing the first sensor to communicate two-bit messages instead of one-bit messages could significantly improve the performance of a two-sensor network for binary hypothesis testing. One way to view this result is as follows: Multi-bit (soft) decisions are able to transmit more information to the FC for the final decision than a binary (hard) decision would. The difficulty is in figuring out how to best capture and quantize this additional information and this problem is the main topic of our work.

Motivated by the above, the main contribution of this paper is to introduce a numerical methodology for designing an $N$-node tandem network of DMs with arbitrary-valued messages. As in \cite{Ek82,Varsh96}, the objective is to design the decision rules at the DMs so as to minimize the overall average cost of making the last decision under the assumption that the observations are conditionally independent. To this end, we propose person-by-person optimization of each DM. However, to arrive at a tractable performance metric for the design (optimization) of each individual DM we design each DM jointly with the FC (fusion center), i.e., the DM is optimized under the assumption that the FC always employs the (optimal) maximum a-posteriori (MAP) rule applied to whatever input it receives. This obviates the need for the number of messages at the output of the DM to be equal to the number of hypotheses, making the proposed method more generally applicable than prior work. Each DM is then also (internally) optimized with respect to the so-obtained metric using a person-by-person method applied to the individual input to output assignments. We finally show that the proposed algorithm is computationally efficient; its complexity per iteration over all DMs is linear in the number of DMs, i.e., the complexity per DM and person-by-person iteration is constant. This is achieved though the novel introduction of an equivalent, restricted, problem formulation for the individual optimization of each DM, and though an efficient recursive computation of the quantities of the equivalent model. Although the proposed design is not globally optimal, because the descent algorithm provided by person-by-person optimization can only be generally guaranteed to converge towards a local optimum for non-convex problems, we show good performance with respect to the few existing benchmark solutions through numerical examples.

The outline of this paper is as follows. In Section \ref{sec:prob} we describe the structure of the tandem network and formulate the problem. In Section \ref{sec:restricted} we introduce the restricted network model, describe how it can be connected to the tandem network, and present the proposed design method. Numerical examples are given in Section \ref{sec:sim} and Section \ref{sec:conc} concludes the paper.
\section{Problem Statement}\label{sec:prob}

We consider a Bayesian decentralized hypothesis testing system with $N$ sensors in a tandem network as shown in Fig.~\ref{fig:tandem}. The sensors, or decision makers (DMs), observe the same phenomenon $H$. DM $l$, using its own observation $x_l\in \mathcal{X}_l$ and the output $u_{l-1}\in \mathcal{M}_{l-1}$ of its predecessor makes a decision $u_l\in\mathcal{M}_l$ and sends it to its successor DM $(l+1)$. The exception to this rule is DM $1$ which using only its own observation $x_1\in\mathcal{X}_1$ makes a decision $u_1\in\mathcal{M}_1$. Throughout this work, the set of possible observations $\mathcal{X}_l$ and the set of possible message\revise{s} $\mathcal{M}_l$ are assumed to be discrete for $l=1,\ldots,N$. Although we restrict our attention to discrete observation spaces, $\mathcal{X}_l$ could be used to approximate observations in a continuous space using fine-grained binning as in \cite{LLG90,Alla14}, where  each bin, or interval, in the continuous observation space can then be represented by an index $x_l$ from the discrete set $\mathcal{X}_l$.

The channel between DM $l$ and its successor DM $(l+1)$ is an error-free and rate-constrained channel of rate $R_l = \log_2 \Vert\mathcal{M}_l\Vert$ bits where $\Vert\mathcal{M}_l\Vert$ denotes the cardinality of $\mathcal{M}_l$. DM $l$ ($l>1$) can be viewed as a quantizer that maps its input vector $(x_l, u_{l-1})$ to an output value (message) $u_l$ using a decision function $\gamma_l: \mathcal{X}_l\times \mathcal{M}_{l-1}\to \mathcal{M}_l$, i.e.,
\begin{equation*}\gamma_l(x_l, u_{l-1})=u_{l}\quad l= 2, \ldots, N\,.\end{equation*} DM $1$ only uses its direct observation $x_1$ to make the decision $u_1$ using a decision function $\gamma_1: \mathcal{X}_1\to \mathcal{M}_1$, i.e., \begin{equation*}\gamma_1(x_1)=u_{1}\,.\end{equation*} Each decision function $\gamma_l$ can also be viewed as an index assignment which assigns an index $u_l$ to each input vector $(x_l, u_{l-1})$ for $l=2,\ldots,N$ or $(x_1)$. DM $N$ makes the global decision $u_N$ in favor of one of the hypotheses.  Without loss of generality we assume that the output message of DM $l$ is from the set $\mathcal{M}_l=\{1, 2, \ldots, 2^{R_l}\}$, while the output message of DM $N$ (the fusion center) is from the set $\mathcal{M}_N=\{1, 2, \ldots, M\}$ for an $M$-ary hypothesis testing problem. We interchangeably use the terms ``message'' and ``index'' for a DM output. We also use both ``DM $N$'' and ``FC'' for the last decision maker, which is also the fusion center of the network.

We assume that the observations at the DMs, conditioned on the hypothesis, are independent, which implies that $x_l$ and $u_{l-1}$, conditioned on the hypothesis, are independent. We also assume that the observation $x_l$ of DM $l$ is a random variable with known conditional probability mass functions (PMF) $P(x_l\vert H_j)$, $j=1, 2, \ldots, M$.

In this paper, as in \cite{Ek82,Varsh96} referred to in the introduction, the objective is to design the tandem network by designing $\gamma_l$ for $l=1,\ldots,N$ in such a way that the global error probability (the error probability of DM $N$) is minimized. We use the person-by-person methodology to numerically derive a decision function at a given DM, under the assumption that all other DMs have already been designed and remain fixed. However in contrast to \cite{Ek82,Varsh96} we treat the FC in a different way than the other DMs: the FC function $\gamma_N$ is always updated together with the DM function $\gamma_l$ currently being optimized, where $l=1,\ldots,N-1$.

For a fixed set of decision functions $\gamma_1$ to $\gamma_{N-1}$, the optimal decision rule for the FC is the maximum a-posteriori (MAP) rule. For this reason, and since the MAP rule allows for a tractable implementation in a single sensor scenario, we will assume that the FC always uses the MAP rule in order to make the global decision $u_N$, given its input $z\triangleq(x_N, u_{N-1})$.
Given $z$, the FC thus decides on $H_{\hat{m}}$ if \cite{Lap09}
\begin{equation}
\pi_{\hat{m}} P(z\vert H_{\hat{m}})=\max_{j}\big\{\pi_{j}P(z\vert H_j)\big\}
\label{eq:FCrule}
\end{equation}
where, $\pi_{j}\triangleq P(H_{j})$ is the a-prior probability of hypothesis $H_j$ and where $j \in \{1, 2, \ldots, M\}$ for the $M$-ary hypothesis testing problem. The expected minimal error probability in estimating $H$ given an observation $z$ from the complete observation set $\mathcal{Z} \triangleq \mathcal{X}_{N} \times \mathcal{M}_{N-1}$ is \cite{Fed94}
\begin{equation}
P_\mathrm{E}=1-\sum_{z\in \mathcal{Z}} \max_{j} \big\{\pi_{j}P(z\vert H_{j})\big\}.
\label{eq:Pe}
\end{equation}
Our objective is to derive decision functions of DM $1$ through $N-1$ that attempts to minimize the expression in \eqref{eq:Pe}, so as to minimize the global error probability.

Letting the FC use the MAP rule in \eqref{eq:FCrule} implies that it always makes the optimum decision based on its input $z$. However, it should be noted that the minimum achievable error probability expression in \eqref{eq:Pe} can be compactly expressed without \emph{explicitly} expressing the FC rule, thus making it a suitable design criterion for the other \revise{DMs}. The criterion only depends on the conditional distributions $P(z|H_j)$ of the FC input and the a-prior probabilities of the hypotheses. One way to view \eqref{eq:Pe} is as a measure of the amount of useful information that is delivered to the FC by the messages from the prior DMs and the FC's own observation.

This view is reminiscent of Longo \emph{et. al}'s design method \cite{LLG90} for parallel networks for binary hypothesis testing where in place of the error probability, they used the Bhattacharyya distance (or equivalently the Bhattacharyya coefficient) applied to the FC input as a performance metric for design of DMs. They designed each DM in a person-by-person manner in such a way that the Bhattacharyya distance at the FC was locally increased at each step. Despite claims to the contrary in \cite{LLG90}, we recently demonstrated in \cite{Alla14} that the same approach could be used in the design of parallel networks with the minimum probability of error expression in place of the Bhattacharyya distance. The minimum probability \revise{of} error design metric also has the added benefit that it extends naturally to $M$-ary hypothesis testing, although this was not discussed in detail in \cite{Alla14}. The optimization through the restricted model introduced next is key to making this approach computationally feasible for long tandem networks.

\section{DM Design Through a Restricted Model}\label{sec:restricted}
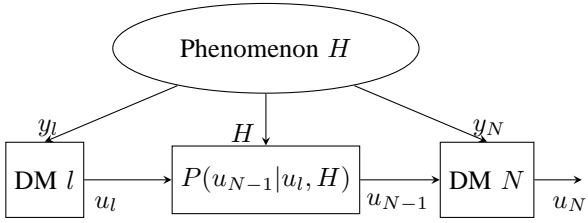
\begin{figure}[t]
\begin{center}
\begin{tikzpicture}[align=center,scale=0.7,>=stealth] 
\node (DM1) at (-2.2,3) [rectangle,minimum size=1.0cm,draw] {DM $l$};
\node (DM2) at (2,3) [rectangle,minimum size=.90cm,draw] {$P(u_{N-1} \vert u_l, H)$};
\node (DMN) at (6.2,3) [rectangle,minimum size=1.0cm,draw] {DM $N$};
\node (PH) at (2,5.5) [ellipse,inner sep=3mm,draw] {Phenomenon $H$};
\draw [->] (DM1.east) -- (DM2) node  [near start,below,inner sep=6pt] {$u_l$};
\draw [->] (DM2.east) -- (DMN) node [near start,below,inner sep=6pt] {}; \node at (4.5,2.57) {$u_{N-1}$};
\draw [->] (DMN.east) --(8,3)   node [near end,below,inner sep=6pt] {$u_N$};
\draw [->] (PH) -- (DM1.north) node [near end,left,inner sep=7pt] {$y_l$};
\draw [->] (PH) -- (DM2.north) node [near end,left,inner sep=4pt] {$H$};
\draw [->] (PH) -- (DMN.north) node [near end,right,inner sep=7pt] {$y_{N}$};
\end{tikzpicture} 
\caption{Restricted model for the tandem network.}
\label{fig:restricted}
\end{center}
\end{figure}

In this section we will show that under the person-by-person methodology, the design of each DM in the tandem network shown in Fig.~\ref{fig:tandem} is analogous to the design of a DM (labelled DM $l$ for notational consistency) in a \emph{restricted} model as shown in Fig.~\ref{fig:restricted}, where DM $N$ in both networks use the MAP rule [cf. \eqref{eq:FCrule}] as the fusion function. Then, using the restricted model, we introduce a computationally efficient algorithm for the design of the DMs.

\subsection{Formation of the Restricted Model}
Consider a distributed system with two DMs as shown in Fig.~\ref{fig:restricted} where each DM has an observation from a discrete observation space $\mathcal{Y}_i$, i.e., $y_i\in \mathcal{Y}_i$, $i=l, N$. DM $l$, using its observation $y_l$, produces a message $u_l$ from the discrete index space $\mathcal{M}_l$ and sends this message to DM $N$ through a discrete channel. DM $N$, as FC of the network, using the received message $u_{N-1} \in \mathcal{M}_{N-1}$ and its own observation $y_N$, makes the global decision $u_N\in \{1, 2, \ldots, M\}$ for an $M$-ary hypothesis testing problem. The channel between the DMs is a discrete channel which maps the index $u_l$ to $u_{N-1}$ with a known transition probability $P(u_{N-1}\vert u_l, H_j)$ which depends on the hypothesis $H$.

Under the person-by-person methodology, the design of DM $l$ ($1\leq l <N $) in the original tandem network of Fig.~\ref{fig:tandem} is analogous to the design of DM $l$ in a particular instance of the restricted model in Fig.~\ref{fig:restricted}. To see this, let
\begin{equation} y_l \triangleq \left\{ \begin{array}{ll}
         x_l & \mbox{if $l=1$}\\
         (x_l, u_{l-1})  & \mbox{if $1<l<N$}  \, , \end{array} \right.
\label{eq:y1def} \end{equation}
be the complete observation of DM $l$ -- combining the direct observation of DM $l$ in the original network and the input from DM ($l-1$) -- and let $y_N \triangleq  x_N$. The conditional PMFs of the inputs to DM $l$ and DM $N$ are due to the independence of $x_l$ and $u_{l-1}$ given by
\begin{equation}
P_j(y_l) = \left\{ \begin{array}{ll}
         P_j(x_l) & \mbox{if $l=1$}\\
         P_j(x_l) P_j(u_{l-1})  & \mbox{if $1<l<N$} ,\end{array} \right.
\label{eq:Py1def}
\end{equation}
\begin{equation*}P_j(y_N) = P_j(x_N).\hspace{40mm}\end{equation*}
The transition probability $P(u_{N-1}\vert u_l, H_j)$ is simply the transition probability from $u_l$ to $u_{N-1}$ in the original network.  The key point is that under the person-by-person design methodology when jointly designing DM $l$ and DM $N$, DM $1$ to DM $(l-1)$ and DM $(l+1)$ to DM $(N-1)$ remain fixed and so does therefore also $P_j(y_l)$ and $P(u_{N-1}\vert u_l, H_j)$. Thus, $P_j(y_l)$ and $P(u_{N-1}\vert u_l, H_j)$ together with the structure of the restricted model in Fig.~\ref{fig:restricted} fully capture all important aspects of the joint design problem for DM $l$ and DM $N$.  In what follows, we will show how to obtain $P_j(y_l)$ [or rather $P_j(u_{l-1})$] and $P(u_{N-1}\vert u_l, H_j)$ in a computationally efficient manner, and how to extend this into an iterative algorithm for the design of the original tandem network.

To this end, consider an arbitrary DM in the original tandem network of Fig.~\ref{fig:tandem}, say, DM $k$. Conditioned on hypothesis $H_j$ each input index $u_{k-1}\in \mathcal{M}_{k-1}$ is mapped to the output index $u_k\in \mathcal{M}_{k}$ with a probability $P_j(u_{k}\vert u_{k-1})\triangleq P(u_{k}\vert u_{k-1}, H_j)$ given by
\begin{equation}
\begin{split}
P_j(u_{k}\vert u_{k-1})
&=P_j\left(\gamma_k(x_k, u_{k-1})\vert u_{k-1}\right)\\
&=\sum_{x_k \in \gamma_k^{-1}(u_{k-1}, u_k)}P_j(x_k) \, ,
\label{eq:inversefun}
\end{split}
\end{equation}
where $P_j(x_k)\triangleq P(x_k\vert H_j)$ is the conditional PMF of $x_k$, and where $\gamma_k^{-1}(u_{k-1}, u_k)$ is the set of observations $x_k$ that satisfy $\gamma_k(x_k, u_{k-1})=u_k$. DM $k$ has a Markovian behavior in the sense that, conditioned on the hypothesis and its input message $u_{k-1}$, the output message $u_k$ depends only upon the direct observation $x_k$, and not the sequence of preceding messages $u_1,\ldots,u_{k-2}$ in the network. The set of DM decisions $u_1,\ldots,u_{N-1}$ thus form a Markov chain, and the probability transitions for this Markov chain can be found using \eqref{eq:inversefun}. The transition probability matrix of DM $k$, conditioned on hypothesis $H_j$, is denoted by ${\bf{P}}^k_j$, has size $\Vert\mathcal{M}_k\Vert\times \Vert\mathcal{M}_{k-1}\Vert$, and an $(m, n)$th entry (by definition) given by ${\bf{P}}^k_j(m, n)\triangleq P_j(u_{k}=m\vert u_{k-1}=n)$ \cite{nor98}.

The Markov property implies that the transition probability from $u_l$ to $u_{N-1}$ in the original tandem network is given by

\begin{align}
P_j(u_{N-1}\vert u_l)
&=\sum_{u_{l+1}}\ldots \sum_{u_{N-2}}P_j(u_{N-1}, u_{N-2}, \ldots, u_{l+1}\vert u_l) \nonumber \\
&=\sum_{u_{l+1}}\ldots \sum_{u_{N-2}}\quad\prod_{i=l+1}^{N-1}P_j(u_{i}\vert u_{i-1}, \ldots, u_l) \nonumber \\
&=\sum_{u_{l+1}}\ldots \sum_{u_{N-2}}\quad\prod_{i=l+1}^{N-1}P_j(u_{i}\vert u_{i-1}) \,. \label{eq:markov1}
\end{align}
Equivalently, in matrix form if we define ${\bf{P}}^{l\to N-1}_j(m, n)\triangleq P_j(u_{N-1}=m\vert u_{l}=n)$, \eqref{eq:markov1} implies
\begin{equation}
{\bf{P}}^{l\to N-1}_j={\bf{P}}^{N-1}_j \times \ldots \times {\bf{P}}^{l+2}_j \times {\bf{P}}^{l+1}_j.
\label{eq:markovmul}
\end{equation}
Thus, using \eqref{eq:markovmul} we can replace all the DMs between DMs $l$ and $N$ by a single hypothesis dependent transition probability given by ${\bf{P}}^{l\to N-1}_j$ when designing DM $l$. Once the transition probability matrix ${\bf{P}}^{l\to N-1}_j$ is found, the probability masses of the messages of DM $N-1$, $P_j(u_{N-1})$, can be easily found from the probability $P_j(u_l)$ of the messages of DM $l$. The complete set of transition probability matrices ${\bf{P}}^{l\to N-1}_j, 1\leq l \leq N-2$ can also be found efficiently (with linear complexity in $N$ per iteration over all DMs) by a recursion with decreasing index $l$, by noting that \eqref{eq:markovmul} implies
\begin{equation}
{\bf{P}}^{l\to N-1}_j\triangleq{\bf{P}}^{l+1\to N-1}_j \times {\bf{P}}^{l+1}_j,
\label{eq:recurmarkov}
\end{equation}
where ${\bf{P}}^{N-1\to N-1}_j={\bf{I}}_{\Vert \mathcal{M}_{N-1}\Vert}$  by definition, and then stored for the forward design of $\gamma_l$ for $l=1,\ldots,N-1$ in one pass of the iterative design algorithm.

By defining the probability mass vector of the messages at the output of DM $k$ as
\begin{equation}
{\bf{q}}^k_j\triangleq \big[P_j(u_k=1), \ldots, P_j(u_k=\Vert \mathcal{M}_{k}\Vert )\big]^T,
\label{eq:qvec}
\end{equation}
the Markov chain property implies \cite{nor98}
\begin{equation}
{\bf{q}}^{N-1}_j={\bf{P}}^{l\to N-1}_j\times {\bf{q}}^{l}_j \, .
\label{eq:relation0}
\end{equation}
Each element of ${\bf{q}}^l_j$ can given $\gamma_l$ (in principle) be found as
\begin{equation}
P_j(u_l)=\sum_{y_l\in \gamma^{-1}_l(u_l)}P_j(y_l), \nonumber
\end{equation}
where $P_j(y_l)$ is defined in \eqref{eq:Py1def}. When $l=1$, $P_j(y_l)$
is simply equal to $P_j(x_1)$ where $x_1$ is the first direct observation in the original network [cf.\ \eqref{eq:inversefun}], while $P_j(y_l)$ for $l > 1$ also depends on $P_j(u_{l-1})$, or equivalently, ${\bf{q}}^{l-1}_j$ for $j=1,\ldots,M$. The latter probability mass vector can however also be obtained recursively by noting that ${\bf{q}}^k_j = {\bf{P}}^{k}_j\times {\bf{q}}^{k-1}_j$ for $k=2,\ldots,l-1$ and that (for $k=1$)

\begin{algorithm}[t]
\caption{Algorithm for designing the DMs in the tandem network Fig.~\ref{fig:tandem}}
\label{alg:tandem}
\begin{algorithmic}[1]
\State \textbf{Input:} Initialized $\gamma_{l}$,\, $l=1, \ldots, N-1$, Iterations $K$
\State \textbf{Output:} Updated $\gamma_{l}$, \, $l=1, \ldots, N-1$
\Initialize{${\bf{P}}^{N-1\to N-1}_j \triangleq {\bf{I}}_{\Vert \mathcal{M}_{N-1}\Vert}$, \, $j=1,\ldots,M$}
\For {$k=1 : K$}
  \For {$l=N-2 : 1$}
\State {find ${\bf{P}}^{l+1}_j$ using \eqref{eq:inversefun}, \, $j=1,\ldots,M$}
\State {${\bf{P}}^{l\to N-1}_j \gets {\bf{P}}^{l+1\to N-1}_j \times {\bf{P}}^{l+1}_j$, \, $j=1,\ldots,M$}
  \EndFor
  \For {$l=1 : N-1$}
\State{optimize $\gamma_l$ using restricted model (cf.\ Alg.\ 2)}
\If {$l=1$}
\State {find ${\bf{q}}^{l}_j$ using \eqref{eq:q1}, \, $j=1,\ldots,M$}
\Else
\State {update ${\bf{P}}^{l}_j$, \, $j=1,\ldots,M$}
\State {${\bf{q}}^{l}_j \gets {\bf{P}}^{l}_j \times {\bf{q}}^{l-1}_j$, \, $j=1,\ldots,M$}
\EndIf
  \EndFor
\EndFor
\end{algorithmic} 
\end{algorithm}
\begin{equation}
{\bf{q}}^{1}_j(m) = \sum_{x_1\in \gamma^{-1}_1(m)}P_j(x_1)
\label{eq:q1}\end{equation}
where $1\leq m \leq \Vert \mathcal{M}_{1}\Vert$. Inserting $P_j(u_{l-1})$ into \eqref{eq:Py1def} gives $P_j(y_l)$ which together with $P_j(u_{k}\vert u_{k-1})$ completely defines the restricted model for the design of DM $l$.

The minimum error probability of a given decision function $\gamma_l$ under MAP decoding at the FC can thus be calculated by calculating $P_j(u_l \vert u_{l-1})$ using \eqref{eq:inversefun}, forming ${\bf{P}}^l_j$, and computing ${\bf{q}}^{N-1}_j = {\bf{P}}^{l\to N-1}_j \times {\bf{P}}^l_j \times {\bf{q}}^{l-1}_j$ which yields $P_j(u_{N-1})$ for $j=1,\ldots,M$; and then applying \eqref{eq:Pe} with $P(z \vert H_j) = P_j(x_N) P_j(u_{N-1})$. This, in principle, allows for optimizing $\gamma_l$ with respect to the global error probability. Note here that both ${\bf{P}}^{l\to N-1}_j$ and ${\bf{q}}^{l-1}_j$ are considered fixed (and precomputed) when designing DM $l$.

Algorithm \ref{alg:tandem} summarizes the overall proposed design procedure of the tandem network, in which for the design of each DM a restricted model should be formed. In each cycle of the optimization, the DMs -- from DM $1$ to DM $(N-1)$ -- are updated one-by-one jointly with DM $N$. After updating DM $l$ its conditional transition probability matrices ${\bf{P}}^{l}_j$ are updated for $j=1,\ldots,M$ and after each cycle the algorithm does another cycle until a given stopping condition is fulfilled (e.g., maximum number of iterations as illustrated in the pseudo-code). The \revise{a}lgorithm then terminates and the last set of decision functions is the final design.

It should be noted that the order in which the optimization of each individual DM is done, \revise{i.e., the order in which Algorithm \ref{alg:restricted} is applied to the set of DMs in the design phase of the network}, as well \revise{how each DM is initialized}, may potentially have an impact on the overall performance of the designed network. This is a consequence of the fact that the proposed person-by-person method is a greedy descent method that only (provably) provides convergence to a local optima. The order could also potentially affect the convergence rate of the design algorithm, and since the performance depends on the order of optimization, the order could in principle be optimized. It is however not clear how this would be done in practice, i.e., what method could be used to determine a good order in a computationally tractable way. This said, the simulation results (in Section \ref{sec:sim}) show that the performance obtained for the proposed ordering, i.e., when the DMs are optimized in a linear order from DM $1$ to DM $N-1$, yields good performance in the few cases where the optimal solution is known, and we have not found any other ordering policies that outperform the one proposed in Algorithm \ref{alg:tandem}. The proposed linear design order is also an essential part of the strategy used to achieve a linear complexity per iteration. \revise{Finally, note here that it is only the order of the optimization of each DM function that is discussed above. The order in which each DM processes its measurement is always fixed as per Fig.~\ref{fig:tandem}, regardless of how the optimization is carried out in the design phase.}

Algorithm \ref{alg:tandem} shows how, regardless of network size, each DM in a tandem network can be designed using the restricted model with a fixed computational burden. Once an explicit design method for the design of the DMs in the restricted model is found, it can be used for the design of a tandem network with arbitrary size $N$, at an overall complexity that grows only linearly in $N$ per iteration between line 4 and line 18 of Algorithm \ref{alg:tandem}. In the next subsection we will introduce a suboptimal, but computationally efficient, method for the design of the DMs in the restricted model.

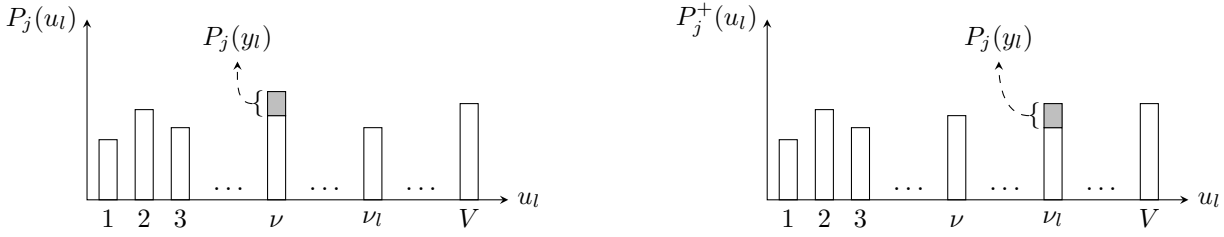
\begin{figure*}[!t]
\begin{minipage}[b]{0.45\linewidth}
\begin{tikzpicture}[align=center,scale=0.8,>=stealth] 
\draw [<->] (0,3) -- (0,0) -- (7,0); \node [left] at (0,3) {$P_j(u_l)$};\node [right] at (7,0) {$u_l$};
\draw (0.2, 0) rectangle (0.5, 1);\node [above] at (.35,-.6) {$1$};
\draw (0.8, 0) rectangle (1.1, 1.5);\node [above] at (.95,-.6) {$2$};
\draw (1.4, 0) rectangle (1.7, 1.2);\node [above] at (1.55,-.6) {$3$};
\node [above] at (2.35,0) {$\ldots$};
\draw (3, 0) rectangle (3.3, 1.4);\node [above] at (3.15,-.6) {$\nu$};
\node [above] at (3.95,0) {$\ldots$};
\draw (4.6, 0) rectangle (4.9, 1.2);\node [above] at (4.75,-.6) {$\nu_l$};
\node [above] at (5.55,0) {$\ldots$};
\draw (6.2, 0) rectangle (6.5, 1.6);\node [above] at (6.35,-.6) {$V$};
\draw [fill=lightgray] (3, 1.4) rectangle (3.3, 1.8);
\draw [dashed, ->] (2.75,1.6) to [out=180,in=270] (2.5,2.3);\node [above] at (2.5,2.3) {$P_j(y_l)$};
\node [left] at (3.1,1.6) {$\{$};
\end{tikzpicture}
\end{minipage}
\hspace{0.5cm}
\begin{minipage}[b]{0.45\linewidth}
\begin{tikzpicture}[align=center,scale=0.8,>=stealth]
\draw [<->] (0,3) -- (0,0) -- (7,0); \node [left] at (0,3) {$P_j^+(u_l)$};\node [right] at (7,0) {$u_l$};
\draw (0.2, 0) rectangle (0.5, 1);\node [above] at (.35,-.6) {$1$};
\draw (0.8, 0) rectangle (1.1, 1.5);\node [above] at (.95,-.6) {$2$};
\draw (1.4, 0) rectangle (1.7, 1.2);\node [above] at (1.55,-.6) {$3$};
\node [above] at (2.35,0) {$\ldots$};
\draw (3, 0) rectangle (3.3, 1.4);\node [above] at (3.15,-.6) {$\nu$};
\node [above] at (3.95,0) {$\ldots$};
\draw (4.6, 0) rectangle (4.9, 1.2);\node [above] at (4.75,-.6) {$\nu_l$};
\node [above] at (5.55,0) {$\ldots$};
\draw (6.2, 0) rectangle (6.5, 1.6);\node [above] at (6.35,-.6) {$V$};
\draw [fill=lightgray] (4.6, 1.2) rectangle (4.9, 1.6);
\draw [dashed, ->] (4.35,1.4) to [out=180,in=270] (3.85,2.3);\node [above] at (3.85,2.3) {$P_j(y_l)$};
\node [left] at (4.7,1.4) {$\{$};
\end{tikzpicture}
\end{minipage}
\caption{An expression of probability masses for the case $\gamma_l(y_l)=\nu$ (left) and when it changes to $\gamma_l^+(y_l)=\nu_l$ (right).}
\label{fig:schematic}
\end{figure*}
\subsection{Design of DMs in the Restricted Model}
From now on our focus will be on the restricted model and we derive the optimization equations for this model, since as explained above the design of DM $l$ in the original tandem network is analogous to the design of DM $l$ in the restricted model with hypothesis dependent transition probability matrices given by ${\bf{P}}^{l\to N-1}_j$ for $j = 1,\ldots, M$. The minimal expected error probability of the restricted model, obtained by MAP decoding at DM $N$, is given by [cf.\ \eqref{eq:Pe}]
\begin{equation}
P_\mathrm{E}=1-\sum_{y_N\in \mathcal{Y}_N}\hspace{2mm}\sum_{u_{N-1}\in \mathcal{M}_{N-1}} \max_{j} \left\{P_j(y_N)P_j(u_{N-1})\pi_j\right\}.
\label{eq:restPe}
\end{equation}

To find the index assignment of each input of DM $l$ that minimizes the global error probability is a combinatorial problem. The total number of possible mappings is given by $\Vert \mathcal{M}_l\Vert^{\Vert \mathcal{Y}_l\Vert}$, where $\Vert \mathcal{Y}_l\Vert=\Vert\mathcal{X}_l\Vert\Vert\mathcal{M}_{l-1}\Vert$, which makes brute force optimal solutions computationally infeasible for any reasonably sized problems. In order to arrive at a computationally efficient procedure, we propose in the following a simple, but suboptimal, method for the design of a particular DM. To do this, we again adopt person-by-person optimization, but now within each individual DM. In other words, the index assignment is done in a person-by-person manner in terms of the input set; an index is assigned to a specific input, while the assigned indices to the other inputs are fixed. Then the optimization formulation for the design of DM $l$ is given as
\begin{equation}\begin{split}
\gamma_{l}^{+}&(y_l)=\\
&\arg \max_{\nu_l \in \mathcal{M}_{l}}\sum_{y_N\in \mathcal{Y}_N}\sum_{u_{N-1}\in \mathcal{M}_{N-1}} \max_{j} \left\{P_j(y_N)P_j(u_{N-1})\pi_j\right\},
\label{eq:optimizer}
\end{split}\end{equation}
where the index assignment $\nu_l$ can change the probability masses in the vector ${\bf{q}}^l_{j}$ [cf. \eqref{eq:qvec}] which consequently affects the PMFs $P_j(u_{N-1})$ through the transition probability matrix ${\bf{P}}^{l\to N-1}_j$ according to \eqref{eq:relation0}. The optimizer of \eqref{eq:optimizer} is found by searching over all possible indices $\nu_l \in \mathcal{M}_{l}$ and for every input $y_l\in \mathcal{Y}_l$. Now let
\begin{equation}
{\bf{P}}^{l\to N-1}_j \triangleq\big[{\bf{r}}_{j,1}\vert {\bf{r}}_{j,2}\vert \ldots \vert {\bf{r}}_{j,\tilde{V}}\big]^T,
\end{equation}
where $\tilde{V}\triangleq \Vert \mathcal{M}_{N-1}\Vert$, and where ${\bf{r}}_{j,m}$ is a column-vector containing the elements of the $m$th row of ${\bf{P}}^{l\to N-1}_j$. Then \eqref{eq:relation0} implies that the $m$th element of ${\bf{q}}_j^{N-1}$, or equivalently $P_j(u_{N-1}=m)$, is found by
\begin{equation}\begin{split}
P_j(u_{N-1}=m)
&= {\bf{r}}_{j,m}^T\times {\bf{q}}_j^{l}\\
&=\big\langle {\bf{q}}_j^{l}, {\bf{r}}_{j,m} \big\rangle,
\label{eq:relation}
\end{split}\end{equation}
where $\langle {\bf{a}},{\bf{b}}\rangle$ is the inner product of the vectors ${\bf{a}}$ and ${\bf{b}}$. Using \eqref{eq:relation} the optimizer \eqref{eq:optimizer} is then written as
\begin{equation}\begin{split}
\gamma_{l}^{+}&(y_l)=\arg \max_{\nu_l \in \mathcal{M}_{l}}\\
&\sum_{y_N\in \mathcal{Y}_N}\sum_{u_{N-1}\in \mathcal{M}_{N-1}} \max_{j}\left \{ \pi_jP_j(y_N)\big\langle {\bf{q}}^{l,(y_l,\nu_l)}_j, {\bf{r}}_{j,u_{N-1}} \big\rangle\right\},
\label{eq:optimizer1}
\end{split}\end{equation}
where we use the superscript $(y_l,\nu_l)$ for the vector ${\bf{q}}^{l,(y_l,\nu_l)}_j$ to emphasize that the assigned index to $y_l$ is $\nu_l$, i.e., $\gamma^+_1(y_l)=\nu_l$. In a shorthand notation,
\begin{equation}
{\bf{q}}^{l,(y_l,\nu_l)}_j\triangleq \big[P_j(u_l=1), \ldots, P_j(u_l=V )\big]^T \Bigg\vert_{\gamma^+_1(y_l)=\nu_l},
\end{equation}
where $V\triangleq\Vert \mathcal{M}_l \Vert $.

It should be noted that after changing the decision function for an input $y_l$, all the conditional probability masses (or equivalently ${\bf{q}}^{l,(y_l,\nu_l)}_j$) need to be calculated which has the potential to make the algorithm difficult to implement for larger rates. However, in the following we will show that only a couple of probability masses in each vector ${{\bf{q}}^{l,(y_l,\nu_l)}_j}$ needs to be updated while the other probability masses remain fixed. Furthermore, we will propose an iterative algorithm for the design of DM $l$ in the restricted model.

To this end, assume now that the decision function for a specific input $y_l\in \mathcal{Y}_l$ evaluates to $\nu\in \mathcal{M}_v$, i.e., $\gamma_l(y_l)=\nu$, and the corresponding conditional PMFs are $P_j(u_l)$, $u_l= 1, 2, \ldots, V$, where the vector ${\bf{q}}^{l,(y_l,\nu)}_j$ is defined as
\begin{equation*}{\bf{q}}^{l,(y_l,\nu)}_j\triangleq \big[P_j(u_l=1), \ldots, P_j(u_l=V )\big]^T\Bigg\vert_{\gamma_l(y_l)=\nu}.
\end{equation*}
Then, the conditional PMFs when $u_l=\nu$ are
\begin{equation}\begin{split}
P_j(u_l=\nu)&=\sum_{y\in \mathcal{Y}_l(\nu)}P_j(y)\\
&= \sum_{\substack{y\in \mathcal{Y}_l(\nu)\\ y\neq y_l}} P_j(y)+P_j(y_l), \nonumber
\end{split}\end{equation}
where $\mathcal{Y}_l(\nu)$ is the set of all inputs $y$ which gives $\gamma_l(y)=\nu$ (including $y_l$).
Assume now that the assigned index to input $y_l$ changes to $\nu_l \in \mathcal{M}_v$ or equivalently $\gamma_l^+(y_l)=\nu_l$. Then $y_l$ does not belong to $\mathcal{Y}_l(\nu)$ anymore (it belongs to $\mathcal{Y}_l(\nu_l)$) and the new conditional PMFs when $u_l=\nu$ and $u_l=\nu_l$ are
\begin{equation}\begin{split}
P^+_j(u_l=\nu)&=\sum_{\substack{y\in \mathcal{Y}_l(\nu)\\ y\neq y_l}} P_j(y)\\
&=P_j(u_l=\nu)-P_j(y_l),\\
P^+_j(u_l=\nu_l)&= P_j(u_l=\nu_l)+P_j(y_l), \nonumber
\label{eq:update1}
\end{split}\end{equation}
while the other conditional probability masses remain fixed at
\begin{equation}
 P^+_j(u_l\neq\nu, \nu_l)=P_j(u_l\neq\nu, \nu_l). \nonumber
\label{eq:update2}
\end{equation}
Consequently, the vector of probability masses for the new index assignment ${\bf{q}}^{l,(y_l,\nu_l)}_j$ can be found from the old vector ${\bf{q}}^{l,(y_l,\nu)}_j$ using
\begin{equation}
{\bf{q}}^{l,(y_l,\nu_l)}_j = {\bf{q}}^{l,(y_l,\nu)}_j + P_j(y_l)({\bf{e}}_{\nu_l}-{\bf{e}}_{\nu}),
\label{eq:update3}
\end{equation}
where ${\bf{e}}_\nu$ is the $\nu$th basis vector in the $V$-dimensional Euclidean space.

This is illustrated in Fig.~\ref{fig:schematic} which shows how the probability masses change when the index assigned to input $y_l$ changes from $\nu$ to $\nu_l$. Thus after updating the index assigned to each input $y_l$ a couple of conditional PMFs, corresponding to the previous and the new assignment, needs to be updated. In other words, only a couple of conditional probability masses in the vector ${\bf{q}}^{l,(y_l,\nu)}_j$ needs to be modified using \eqref{eq:update3}, while the other probability masses remain fixed.

Consider again the optimizer \eqref{eq:optimizer1} for updating the assigned index to input $y_l$. Assume that the assigned index to input $y_l$ prior to updating it is $\nu$, i.e., $\gamma_1(y_l)=\nu$ and the corresponding vector of probability masses is ${\bf{q}}^{l,(y_l,\nu)}_j$. Using \eqref{eq:update3} the inner product $\big\langle {\bf{q}}^{l,(y_l,\nu_l)}_j, {\bf{r}}_{j,u_{N-1}}\big \rangle$ is written
\begin{equation}\nonumber
\begin{split}
\big\langle {\bf{q}}^{l,(y_l,\nu_l)}_j, &{\bf{r}}_{j,u_{N-1}}\big \rangle\\
&=\big\langle {\bf{q}}^{l,(y_l,\nu)}_j + P_j(y_l)({\bf{e}}_{\nu_l}-{\bf{e}}_{\nu}), {\bf{r}}_{j,u_{N-1}} \big\rangle\\
&=\big\langle {\bf{q}}^{l,(y_l,\nu)}_j, {\bf{r}}_{j,u_{N-1}} \big\rangle +P_j(y_l)\langle ({\bf{e}}_{\nu_l}-{\bf{e}}_{\nu}), {\bf{r}}_{j,u_{N-1}} \big\rangle\\
&=\big\langle {\bf{q}}^{l,(y_l,\nu)}_j, {\bf{r}}_{j,u_{N-1}} \big\rangle + P_j(y_l) \Delta r_{j,u_{N-1}}(\nu,\nu_l),
\end{split}
\end{equation}
where
\begin{equation}
\Delta r_{j,u_{N-1}}(\nu,\nu_l)\triangleq {\bf{r}}_{j,u_{N-1}}(\nu_l)-{\bf{r}}_{j,u_{N-1}}(\nu).
\end{equation}
The optimization problem in \eqref{eq:optimizer1} can be written as
\begin{equation}\begin{split}
\gamma_{l}&^{+}(y_l)=
\arg \max_{\nu_l \in \mathcal{M}_{l}}\sum_{y_N\in \mathcal{Y}_N}\hspace{2mm}\sum_{u_{N-1}\in \mathcal{M}_{N-1}}\max_{j} \Bigg\{\\
&\pi_j P_j(y_N)\bigg[\big\langle {\bf{q}}^{l,(y_l,\nu)}_j, {\bf{r}}_{j,u_{N-1}} \big\rangle +
P_j(y_l)\Delta r_{j,u_{N-1}}(\nu,\nu_l)\bigg]\Bigg\},
\label{eq:optimizer2}
\end{split}\end{equation}
where $\nu$ is the assigned index to input $y_l$ prior to updating it.

The updating rule for the design of DM $l$ in the restricted model is described in Algorithm \ref{alg:restricted}. In this algorithm, after updating all the input indices $y_l$, the conditional PMFs of DM $l$ are updated and the performance improvement (the improvement in error probability $\Delta P_{\mathrm{E}}$) is calculated. If it is greater than a threshold $\eta$ the algorithm does another cycle. Otherwise it terminates and the last index assignment for DM $l$ is the final index assignment.

\begin{algorithm}[t]
\caption{Algorithm for designing DM $l$ in restricted model Fig.~\ref{fig:restricted}}
\label{alg:restricted}
\begin{algorithmic}[1]
\State \textbf{Input:} $\gamma_{l}$, $\eta$, $P_j(y_l)$, $P_j(y_N)$ and $P_j(u_{N-1}\vert u_l)$
\State \textbf{Output:} Updated $\gamma_l$
\State set $\Delta P_{\mathrm{E}}\gets\infty$ and find $P_{\mathrm{E}}$ using \eqref{eq:restPe}
\While{$\Delta P_{\mathrm{E}}>\eta$}
    \For  {$i=1 : \Vert \mathcal{Y}_l \Vert$}
    \State        $y_l \gets \mathcal{Y}_l(i)$
    \State        update the assigned index to input $y_l$ using \eqref{eq:optimizer2}
    \For  {$j=1 : M$}
    \State       update vector ${\bf{q}}_j$ using \eqref{eq:update3}
    \EndFor
    \EndFor
\State find $P^+_{\mathrm{E}}$ using \eqref{eq:restPe} and evaluate $\Delta P_{\mathrm{E}}=P_{\mathrm{E}} - P^+_{\mathrm{E}}$
\State $P_{\mathrm{E}}\gets P^+_{\mathrm{E}}$
\EndWhile
\end{algorithmic} 
\end{algorithm}

In closing, we should mention that the optimizer \eqref{eq:optimizer2}, which is equal to $1-P_{\mathrm{E}}$, used for the design of DMs arranged in tandem, has a close relation to the true error probability at the FC [cf. \eqref{eq:restPe}, \eqref{eq:optimizer}]. While we are updating each DM, we try to minimize the error probability of the network, while the other DMs are kept fixed. The error probability is therefore decreased gradually until it converge to a locally optimal solution. Note that this hold\revise{s} for any chosen termination threshold $\eta$ in Algorithm 2 as any update improves the overall error probability. It can also be shown that Algorithm 2 terminates in a finite number of steps for any $\eta > 0$ as it is a descent algorithm over a finite space, and as it will thus never visit the same potential solution twice.

\subsection{Complexity of the proposed method}
To get a more granular view of the complexity of the proposed method, we find the cost of the proposed numerical method by giving the total number of multiplications required. 
We begin by considering the complexity of one pass of the outer \textbf{for} loop of Algorithm \ref{alg:tandem}, and
for the sake of simplicity, we assume all the channels have equal rates, i.e., $\Vert\mathcal{M}\Vert=\Vert\mathcal{M}_1\Vert=\ldots=\Vert\mathcal{M}_{N-1}\Vert$. Each matrix-matrix multiplication in line $7$ of Algorithm \ref{alg:tandem} needs $\Vert\mathcal{M}\Vert^3$ multiplications and each matrix-vector multiplication in line $15$ requires $\Vert\mathcal{M}\Vert^2$ multiplications. Lines $5-8$ in the first inner {\bf for} loop are executed $(N-2)$ times at a total complexity of $(N-2)M\Vert\mathcal{M}\Vert^3$ multiplications, and lines $9-17$ in the second inner {\bf for} loop are executed $(N-1)$ times at a total complexity of $(N-2)M\Vert\mathcal{M}\Vert^2+(N-1)C_{II}$ multiplications, where $C_\mathrm{II}$ is the complexity of  line $10$ (Algorithm \ref{alg:restricted}). This implies that the total complexity of one pass through Algorithm \ref{alg:tandem}'s outer \textbf{for} loop is
\begin{equation}
C_\mathrm{I}=(N-2)M\Vert\mathcal{M}\Vert^3+(N-2)M\Vert\mathcal{M}\Vert^2+(N-1)C_{II}\,.
\label{eq:c1}\end{equation}
In Algorithm \ref{alg:restricted}, finding $P_\mathrm{E}$ (in lines $3$ and $12$) requires $2M\Vert\mathcal{X}\Vert\Vert\mathcal{M}\Vert$ multiplications, where again for simplicity we assumed $\Vert\mathcal{X}\Vert=\Vert\mathcal{X}_1\Vert=\ldots=\Vert\mathcal{X}_{N}\Vert$. Updating the assigned index to each input $y_l$ in \eqref{eq:optimizer2} requires $M\Vert\mathcal{M}\Vert^2\Vert\mathcal{X}\Vert\left(3+\Vert\mathcal{M}\Vert \right)$ multiplications, and updating the vector ${\bf{q}}_j$ requires one multiplication. Lines $6-10$ of Algorithm are repeated $\Vert \mathcal{Y}_l\Vert=\Vert\mathcal{X}\Vert\Vert\mathcal{M}\Vert$ times on each pass, 
and the whole complexity of Algorithm \ref{alg:restricted} if it carries out $T$ iterations of the {\bf while} loop becomes
\begin{equation}\begin{split}
C_\mathrm{II}=&2(1+T)M\Vert\mathcal{X}\Vert\Vert\mathcal{M}\Vert+\\&3MT\Vert\mathcal{X}\Vert^2\Vert\mathcal{M}\Vert^3+
MT\Vert\mathcal{X}\Vert^2\Vert\mathcal{M}\Vert^4\,.
\label{eq:c2}\end{split}
\end{equation}

After plugging \eqref{eq:c2} into \eqref{eq:c1} and dropping dominated terms, the overall complexity of Algorithm \ref{alg:tandem} is approximately given by
\begin{equation}
C_\mathrm{I}\approx NMT\Vert\mathcal{X}\Vert^2\Vert\mathcal{M}\Vert^4
\label{eq:C1}\end{equation}
multiplications per iteration.

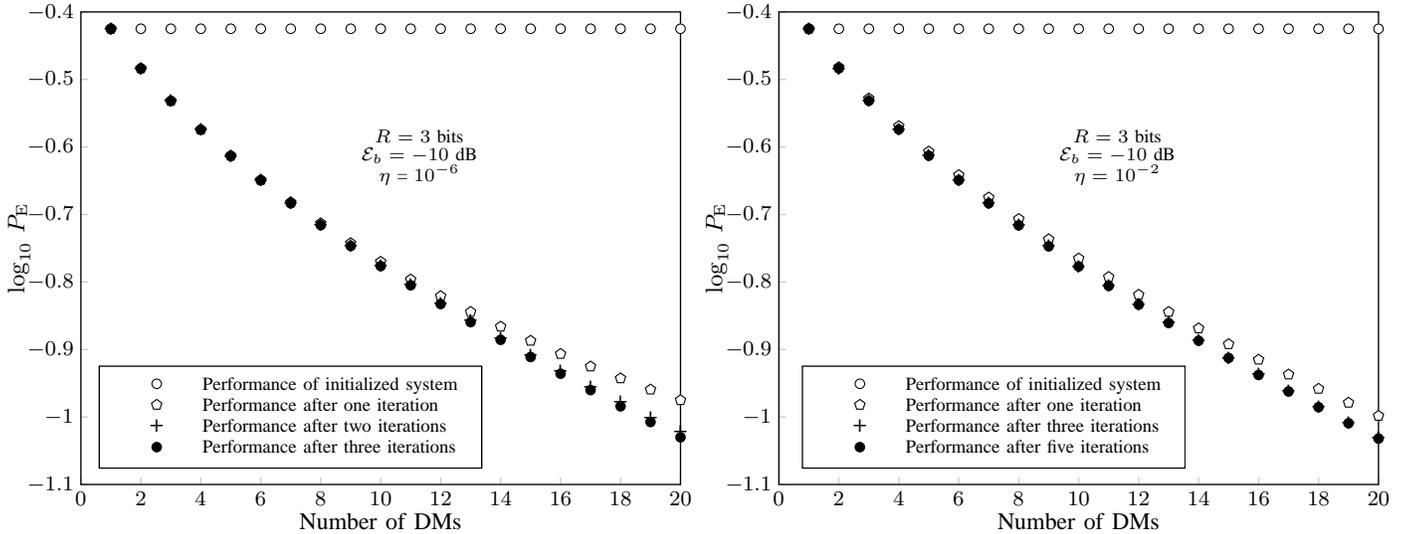
\begin{figure*}[!t]
\begin{minipage}[b]{0.5\linewidth}
\psset{xunit=0.050000\plotwidth,yunit=1.126728\plotwidth}%
\begin{pspicture}(-2.396313,-1.177778)(20.368664,-0.383626)%


\psline[linewidth=\AxesLineWidth,linecolor=GridColor](0.000000,-1.100000)(0.000000,-1.089350)
\psline[linewidth=\AxesLineWidth,linecolor=GridColor](2.000000,-1.100000)(2.000000,-1.089350)
\psline[linewidth=\AxesLineWidth,linecolor=GridColor](4.000000,-1.100000)(4.000000,-1.089350)
\psline[linewidth=\AxesLineWidth,linecolor=GridColor](6.000000,-1.100000)(6.000000,-1.089350)
\psline[linewidth=\AxesLineWidth,linecolor=GridColor](8.000000,-1.100000)(8.000000,-1.089350)
\psline[linewidth=\AxesLineWidth,linecolor=GridColor](10.000000,-1.100000)(10.000000,-1.089350)
\psline[linewidth=\AxesLineWidth,linecolor=GridColor](12.000000,-1.100000)(12.000000,-1.089350)
\psline[linewidth=\AxesLineWidth,linecolor=GridColor](14.000000,-1.100000)(14.000000,-1.089350)
\psline[linewidth=\AxesLineWidth,linecolor=GridColor](16.000000,-1.100000)(16.000000,-1.089350)
\psline[linewidth=\AxesLineWidth,linecolor=GridColor](18.000000,-1.100000)(18.000000,-1.089350)
\psline[linewidth=\AxesLineWidth,linecolor=GridColor](20.000000,-1.100000)(20.000000,-1.089350)
\psline[linewidth=\AxesLineWidth,linecolor=GridColor](0.000000,-1.100000)(0.240000,-1.100000)
\psline[linewidth=\AxesLineWidth,linecolor=GridColor](0.000000,-1.000000)(0.240000,-1.000000)
\psline[linewidth=\AxesLineWidth,linecolor=GridColor](0.000000,-0.900000)(0.240000,-0.900000)
\psline[linewidth=\AxesLineWidth,linecolor=GridColor](0.000000,-0.800000)(0.240000,-0.800000)
\psline[linewidth=\AxesLineWidth,linecolor=GridColor](0.000000,-0.700000)(0.240000,-0.700000)
\psline[linewidth=\AxesLineWidth,linecolor=GridColor](0.000000,-0.600000)(0.240000,-0.600000)
\psline[linewidth=\AxesLineWidth,linecolor=GridColor](0.000000,-0.500000)(0.240000,-0.500000)
\psline[linewidth=\AxesLineWidth,linecolor=GridColor](0.000000,-0.400000)(0.240000,-0.400000)

{ \footnotesize 
\rput[t](0.000000,-1.110650){$0$}
\rput[t](2.000000,-1.110650){$2$}
\rput[t](4.000000,-1.110650){$4$}
\rput[t](6.000000,-1.110650){$6$}
\rput[t](8.000000,-1.110650){$8$}
\rput[t](10.000000,-1.110650){$10$}
\rput[t](12.000000,-1.110650){$12$}
\rput[t](14.000000,-1.110650){$14$}
\rput[t](16.000000,-1.110650){$16$}
\rput[t](18.000000,-1.110650){$18$}
\rput[t](20.000000,-1.110650){$20$}
\rput[r](-0.240000,-1.100000){$-1.1$}
\rput[r](-0.240000,-1.000000){$-1$}
\rput[r](-0.240000,-0.900000){$-0.9$}
\rput[r](-0.240000,-0.800000){$-0.8$}
\rput[r](-0.240000,-0.700000){$-0.7$}
\rput[r](-0.240000,-0.600000){$-0.6$}
\rput[r](-0.240000,-0.500000){$-0.5$}
\rput[r](-0.240000,-0.400000){$-0.4$}
} 

\psframe[linewidth=\AxesLineWidth,dimen=middle](0.000000,-1.100000)(20.000000,-0.400000)

{ \small 
\rput[b](10.000000,-1.177778){
\begin{tabular}{c}
Number of DMs\\
\end{tabular}
}

\rput[t]{90}(-2.5,-0.750000){
\begin{tabular}{c}
$\log_{10}$ $P_\mathrm{E}$\\
\end{tabular}
}
} 

\newrgbcolor{color264.0018}{0  0  0}
\psline[plotstyle=line,linejoin=1,showpoints=true,dotstyle=Bo,dotsize=\MarkerSize,linestyle=none,linewidth=\LineWidth,linecolor=color264.0018]
(1.000000,-0.424928)(2.000000,-0.424900)(3.000000,-0.424900)(4.000000,-0.424900)(5.000000,-0.424900)
(6.000000,-0.424900)(7.000000,-0.424900)(8.000000,-0.424900)(9.000000,-0.424900)(10.000000,-0.424900)
(11.000000,-0.424900)(12.000000,-0.424900)(13.000000,-0.424900)(14.000000,-0.424900)(15.000000,-0.424900)
(16.000000,-0.424900)(17.000000,-0.424900)(18.000000,-0.424900)(19.000000,-0.424900)(20.000000,-0.424900)

\newrgbcolor{color265.0012}{0  0  0}
\psline[plotstyle=line,linejoin=1,showpoints=true,dotstyle=Bpentagon,dotsize=\MarkerSize,linestyle=none,linewidth=\LineWidth,linecolor=color265.0012]
(1.000000,-0.424928)(2.000000,-0.483597)(3.000000,-0.531800)(4.000000,-0.574303)(5.000000,-0.612966)
(6.000000,-0.648784)(7.000000,-0.681937)(8.000000,-0.713095)(9.000000,-0.742561)(10.000000,-0.770062)
(11.000000,-0.796423)(12.000000,-0.821023)(13.000000,-0.844360)(14.000000,-0.866142)(15.000000,-0.887060)
(16.000000,-0.906578)(17.000000,-0.925184)(18.000000,-0.942714)(19.000000,-0.959398)(20.000000,-0.975104)

\newrgbcolor{color289.0013}{0  0  0}
\psline[plotstyle=line,linejoin=1,showpoints=true,dotstyle=B+,dotsize=\MarkerSize,linestyle=none,linewidth=\LineWidth,linecolor=color289.0013]
(1.000000,-0.424928)(2.000000,-0.483597)(3.000000,-0.531800)(4.000000,-0.574466)(5.000000,-0.613144)
(6.000000,-0.649364)(7.000000,-0.683191)(8.000000,-0.715344)(9.000000,-0.745936)(10.000000,-0.775208)
(11.000000,-0.803547)(12.000000,-0.830914)(13.000000,-0.857298)(14.000000,-0.882729)(15.000000,-0.907630)
(16.000000,-0.931814)(17.000000,-0.955068)(18.000000,-0.977984)(19.000000,-1.000435)(20.000000,-1.022276)

\newrgbcolor{color290.0013}{0  0  0}
\psline[plotstyle=line,linejoin=1,showpoints=true,dotstyle=*,dotsize=\MarkerSize,linestyle=none,linewidth=\LineWidth,linecolor=color290.0013]
(1.000000,-0.424928)(2.000000,-0.483597)(3.000000,-0.531800)(4.000000,-0.574466)(5.000000,-0.613323)
(6.000000,-0.649364)(7.000000,-0.683401)(8.000000,-0.715795)(9.000000,-0.746662)(10.000000,-0.776504)
(11.000000,-0.804931)(12.000000,-0.832683)(13.000000,-0.859492)(14.000000,-0.885723)(15.000000,-0.911155)
(16.000000,-0.935917)(17.000000,-0.960189)(18.000000,-0.984221)(19.000000,-1.007446)(20.000000,-1.030118)

{ \scriptsize 
\rput(7,-1.){%
\psshadowbox[framesep=0pt,shadowsize=0pt,linewidth=\AxesLineWidth]{\psframebox*{\begin{tabular}{l}
\Rnode{a1}{\hspace*{0.0ex}} \hspace*{0.7cm} \Rnode{a2}{~~Performance of initialized system} \\
\Rnode{a3}{\hspace*{0.0ex}} \hspace*{0.7cm} \Rnode{a4}{~~Performance after one iteration} \\
\Rnode{a5}{\hspace*{0.0ex}} \hspace*{0.7cm} \Rnode{a6}{~~Performance after two iterations} \\
\Rnode{a7}{\hspace*{0.0ex}} \hspace*{0.7cm} \Rnode{a8}{~~Performance after three iterations} \\
\end{tabular}}
\ncline[linestyle=none,linewidth=\LineWidth,linecolor=color264.0018]{a1}{a2} \ncput{\psdot[dotstyle=Bo,dotsize=\MarkerSize,linecolor=color264.0018]}
\ncline[linestyle=none,linewidth=\LineWidth,linecolor=color265.0012]{a3}{a4} \ncput{\psdot[dotstyle=Bpentagon,dotsize=\MarkerSize,linecolor=color265.0012]}
\ncline[linestyle=none,linewidth=\LineWidth,linecolor=color289.0013]{a5}{a6} \ncput{\psdot[dotstyle=B+,dotsize=\MarkerSize,linecolor=color289.0013]}
\ncline[linestyle=none,linewidth=\LineWidth,linecolor=color290.0013]{a7}{a8} \ncput{\psdot[dotstyle=*,dotsize=\MarkerSize,linecolor=color290.0013]}
}%
}%
} 

{ \scriptsize 
\newrgbcolor{color208.0011}{0  0  0}
\uput{0pt}[0](9.226190,-0.613843){%
\psframebox[framesep=1pt,fillstyle=solid,linestyle=none,linewidth=0.5pt]{\begin{tabular}{@{}c@{}}
$R$ $=$ $3$ bits\\[-0.3ex]
$\mathcal{E}_b$ $=$ $-10$ dB\\[-0.3ex]
$\eta$ = $10^{-6}$\\[-0.3ex]
\end{tabular}}}
} 

\end{pspicture}%
\end{minipage}
\hspace{0.05cm}
\begin{minipage}[b]{0.5\linewidth}
\psset{xunit=0.050000\plotwidth,yunit=1.126728\plotwidth}%
\begin{pspicture}(-2.396313,-1.177778)(20.368664,-0.383626)%


\psline[linewidth=\AxesLineWidth,linecolor=GridColor](0.000000,-1.100000)(0.000000,-1.089350)
\psline[linewidth=\AxesLineWidth,linecolor=GridColor](2.000000,-1.100000)(2.000000,-1.089350)
\psline[linewidth=\AxesLineWidth,linecolor=GridColor](4.000000,-1.100000)(4.000000,-1.089350)
\psline[linewidth=\AxesLineWidth,linecolor=GridColor](6.000000,-1.100000)(6.000000,-1.089350)
\psline[linewidth=\AxesLineWidth,linecolor=GridColor](8.000000,-1.100000)(8.000000,-1.089350)
\psline[linewidth=\AxesLineWidth,linecolor=GridColor](10.000000,-1.100000)(10.000000,-1.089350)
\psline[linewidth=\AxesLineWidth,linecolor=GridColor](12.000000,-1.100000)(12.000000,-1.089350)
\psline[linewidth=\AxesLineWidth,linecolor=GridColor](14.000000,-1.100000)(14.000000,-1.089350)
\psline[linewidth=\AxesLineWidth,linecolor=GridColor](16.000000,-1.100000)(16.000000,-1.089350)
\psline[linewidth=\AxesLineWidth,linecolor=GridColor](18.000000,-1.100000)(18.000000,-1.089350)
\psline[linewidth=\AxesLineWidth,linecolor=GridColor](20.000000,-1.100000)(20.000000,-1.089350)
\psline[linewidth=\AxesLineWidth,linecolor=GridColor](0.000000,-1.100000)(0.240000,-1.100000)
\psline[linewidth=\AxesLineWidth,linecolor=GridColor](0.000000,-1.000000)(0.240000,-1.000000)
\psline[linewidth=\AxesLineWidth,linecolor=GridColor](0.000000,-0.900000)(0.240000,-0.900000)
\psline[linewidth=\AxesLineWidth,linecolor=GridColor](0.000000,-0.800000)(0.240000,-0.800000)
\psline[linewidth=\AxesLineWidth,linecolor=GridColor](0.000000,-0.700000)(0.240000,-0.700000)
\psline[linewidth=\AxesLineWidth,linecolor=GridColor](0.000000,-0.600000)(0.240000,-0.600000)
\psline[linewidth=\AxesLineWidth,linecolor=GridColor](0.000000,-0.500000)(0.240000,-0.500000)
\psline[linewidth=\AxesLineWidth,linecolor=GridColor](0.000000,-0.400000)(0.240000,-0.400000)

{ \footnotesize 
\rput[t](0.000000,-1.110650){$0$}
\rput[t](2.000000,-1.110650){$2$}
\rput[t](4.000000,-1.110650){$4$}
\rput[t](6.000000,-1.110650){$6$}
\rput[t](8.000000,-1.110650){$8$}
\rput[t](10.000000,-1.110650){$10$}
\rput[t](12.000000,-1.110650){$12$}
\rput[t](14.000000,-1.110650){$14$}
\rput[t](16.000000,-1.110650){$16$}
\rput[t](18.000000,-1.110650){$18$}
\rput[t](20.000000,-1.110650){$20$}
\rput[r](-0.240000,-1.100000){$-1.1$}
\rput[r](-0.240000,-1.000000){$-1$}
\rput[r](-0.240000,-0.900000){$-0.9$}
\rput[r](-0.240000,-0.800000){$-0.8$}
\rput[r](-0.240000,-0.700000){$-0.7$}
\rput[r](-0.240000,-0.600000){$-0.6$}
\rput[r](-0.240000,-0.500000){$-0.5$}
\rput[r](-0.240000,-0.400000){$-0.4$}
} 

\psframe[linewidth=\AxesLineWidth,dimen=middle](0.000000,-1.100000)(20.000000,-0.400000)

{ \small 
\rput[b](10.000000,-1.177778){
\begin{tabular}{c}
Number of DMs\\
\end{tabular}
}

\rput[t]{90}(-2.5,-0.750000){
\begin{tabular}{c}
$\log_{10}$ $P_\mathrm{E}$\\
\end{tabular}
}
} 

\newrgbcolor{color620.0027}{0  0  0}
\psline[plotstyle=line,linejoin=1,showpoints=true,dotstyle=Bo,dotsize=\MarkerSize,linestyle=none,linewidth=\LineWidth,linecolor=color620.0027]
(1.000000,-0.424928)(2.000000,-0.424928)(3.000000,-0.424928)(4.000000,-0.424928)(5.000000,-0.424928)
(6.000000,-0.424928)(7.000000,-0.424928)(8.000000,-0.424928)(9.000000,-0.424928)(10.000000,-0.424928)
(11.000000,-0.424928)(12.000000,-0.424928)(13.000000,-0.424928)(14.000000,-0.424928)(15.000000,-0.424928)
(16.000000,-0.424928)(17.000000,-0.424928)(18.000000,-0.424928)(19.000000,-0.424928)(20.000000,-0.424928)

\newrgbcolor{color621.0022}{0  0  0}
\psline[plotstyle=line,linejoin=1,showpoints=true,dotstyle=Bpentagon,dotsize=\MarkerSize,linestyle=none,linewidth=\LineWidth,linecolor=color621.0022]
(1.000000,-0.424928)(2.000000,-0.481881)(3.000000,-0.528122)(4.000000,-0.569119)(5.000000,-0.606776)
(6.000000,-0.641684)(7.000000,-0.674895)(8.000000,-0.706637)(9.000000,-0.736838)(10.000000,-0.765230)
(11.000000,-0.792635)(12.000000,-0.819014)(13.000000,-0.844360)(14.000000,-0.868702)(15.000000,-0.892112)
(16.000000,-0.915066)(17.000000,-0.937042)(18.000000,-0.958213)(19.000000,-0.978811)(20.000000,-0.998266)

\newrgbcolor{color622.0022}{0  0  0}
\psline[plotstyle=line,linejoin=1,showpoints=true,dotstyle=B+,dotsize=\MarkerSize,linestyle=none,linewidth=\LineWidth,linecolor=color622.0022]
(1.000000,-0.424928)(2.000000,-0.483200)(3.000000,-0.531062)(4.000000,-0.573651)(5.000000,-0.612610)
(6.000000,-0.648590)(7.000000,-0.682982)(8.000000,-0.715344)(9.000000,-0.746662)(10.000000,-0.776504)
(11.000000,-0.805208)(12.000000,-0.832978)(13.000000,-0.860121)(14.000000,-0.886391)(15.000000,-0.911864)
(16.000000,-0.936667)(17.000000,-0.960983)(18.000000,-0.984221)(19.000000,-1.007889)(20.000000,-1.031050)

\newrgbcolor{color623.0022}{0  0  0}
\psline[plotstyle=line,linejoin=1,showpoints=true,dotstyle=*,dotsize=\MarkerSize,linestyle=none,linewidth=\LineWidth,linecolor=color623.0022]
(1.000000,-0.424928)(2.000000,-0.483465)(3.000000,-0.531505)(4.000000,-0.573977)(5.000000,-0.612966)
(6.000000,-0.649171)(7.000000,-0.683401)(8.000000,-0.716021)(9.000000,-0.747147)(10.000000,-0.777024)
(11.000000,-0.805763)(12.000000,-0.833570)(13.000000,-0.860751)(14.000000,-0.887060)(15.000000,-0.912574)
(16.000000,-0.937794)(17.000000,-0.962175)(18.000000,-0.985479)(19.000000,-1.009217)(20.000000,-1.031984)

{ \scriptsize 
\rput(7.173072,-1){%
\psshadowbox[framesep=0pt,shadowsize=0pt,linewidth=\AxesLineWidth]{\psframebox*{\begin{tabular}{l}
\Rnode{a1}{\hspace*{0.0ex}} \hspace*{0.7cm} \Rnode{a2}{~~Performance of initialized system} \\
\Rnode{a3}{\hspace*{0.0ex}} \hspace*{0.7cm} \Rnode{a4}{~~Performance after one iteration} \\
\Rnode{a5}{\hspace*{0.0ex}} \hspace*{0.7cm} \Rnode{a6}{~~Performance after three iterations} \\
\Rnode{a7}{\hspace*{0.0ex}} \hspace*{0.7cm} \Rnode{a8}{~~Performance after five iterations} \\
\end{tabular}}
\ncline[linestyle=none,linewidth=\LineWidth,linecolor=color620.0027]{a1}{a2} \ncput{\psdot[dotstyle=Bo,dotsize=\MarkerSize,linecolor=color620.0027]}
\ncline[linestyle=none,linewidth=\LineWidth,linecolor=color621.0022]{a3}{a4} \ncput{\psdot[dotstyle=Bpentagon,dotsize=\MarkerSize,linecolor=color621.0022]}
\ncline[linestyle=none,linewidth=\LineWidth,linecolor=color622.0022]{a5}{a6} \ncput{\psdot[dotstyle=B+,dotsize=\MarkerSize,linecolor=color622.0022]}
\ncline[linestyle=none,linewidth=\LineWidth,linecolor=color623.0022]{a7}{a8} \ncput{\psdot[dotstyle=*,dotsize=\MarkerSize,linecolor=color623.0022]}
}%
}%
} 

{ \scriptsize 
\newrgbcolor{color549.0021}{0  0  0}
\uput{0pt}[0](9.226190,-0.613843){%
\psframebox[framesep=1pt,fillstyle=solid,linestyle=none,linewidth=0.5pt]{\begin{tabular}{@{}c@{}}
$R$ $=$ $3$ bits\\[-0.3ex]
$\mathcal{E}_b$ $=$ $-10$ dB\\[-0.3ex]
$\eta$ $=$ $10^{-2}$\\[-0.3ex]
\end{tabular}}}
} 

\end{pspicture}%
\end{minipage}
\caption{Evolution of error probability of tandem networks with different number of DMs, \revise{for $\eta=10^{-6}$ (left), and for $\eta=10^{-2}$ (right)}.}
\label{fig:conv}
\end{figure*}

In the resulting complexity $C_I$ in \eqref{eq:C1} for one iteration of Algorithm \ref{alg:tandem} the parameter $T$ (the number of iterations that Algorithm \ref{alg:restricted} carries out) depends in some nontrivial way on the parameter $\eta$, and it is not within the scope of this work to characterize the dependence of $T$ on $\eta$ (or more generally on the stopping criterion in Algorithm \ref{alg:restricted}). It is however clear that $T$ is non increasing in (increasing) $\eta$, and upper bounded by some function of $\Vert \mathcal{X}\Vert$ and $\Vert \mathcal{M} \Vert$ independently of $\eta > 0$ as noted before. Any easily provable bound on $T$ is however likely to be too loose to be of much use other than in the theoretical proof of linear complexity per iteration of Algorithm \ref{alg:tandem}.
This said, in our simulations we never observed a value of $T$ above 4 for $\eta = 10^{-6}$.

\section{Simulations}\label{sec:sim}
In this section we present some results illustrating the application of the proposed method in the design of tandem networks. To show the performance of the proposed method, and to enable comparisons with the very few existing closed form design rules, we first consider the case of binary hypothesis testing, i.e., $M=2$. Next, in order to illustrate the benefits of the proposed method in $M$-ary hypothesis testing, we consider the performance of the designed tandem networks in ternary and quaternary hypothesis testing, i.e., $M=3$ and $M=4$. We limit our attention to independent and identically distributed observations $x_i$, $i=1, 2, \ldots, N$ where each real valued observation consists of a known signal in additive white Gaussian noise $\mathcal{N}(0,\sigma^2)$.

\subsection{Binary hypothesis testing} We first consider a binary hypothesis testing problem in which each real valued observation consists of an antipodal signal $\pm a$ in unit-variance ($\sigma^2=1$) additive white Gaussian \revise{noise}. The observation model at each DM is \footnote{It should be mentioned that in this paper the hypothesis set $\{H_1, \ldots, H_M\}$ is used for $M$-ary hypothesis testing, while for $M=2$ we use the hypothesis set $\{H_0,H_1\}$ instead of $\{H_1, H_2\}$ for notational consistency with existing texts on binary hypothesis testing.}
\begin{equation*}
\begin{split}
&H_0: {x_i}={-a} +{n_i}\revise{\,,} \\
&H_1: {x_i}={+a} + {n_i}\,.
\end{split}
\end{equation*}
We also define the per channel signal-to-noise ratio (SNR) for binary hypothesis test as $\mathcal{E}_b\triangleq\vert a\vert^2$, and assume that the hypotheses are equally likely ($\pi_0=\pi_1=0.5$). Furthermore the channel rates are considered to be the same for all links and equal to $R$ which implies the DMs output messages are from the set $\mathcal{M} = \,\{1, 2, \ldots, 2^R\}$.

Although the proposed design method is for discrete observation sets, it can be applied to the continuous real valued observations using fine-grained binning \cite{Alla14}. To do that, the interval $[-a-4, a+4]$ (containing $0.9997$ of the total probability mass for each DM) is represented by $128$ discrete probability masses per hypothesis to form discrete observation sets from the continuous observations, i.e., $\Vert \mathcal{X}_k\Vert=128$.

Fig.~\ref{fig:conv} shows the evolution of the error probability for designed tandem networks with $1 \leq N \leq 20$ DMs after \revise{$K=\{1,2,3\}$} iterations of Algorithm \ref{alg:tandem} for $\eta=10^{-6}$ in Algorithm \ref{alg:restricted}\revise{, and after $K=\{1,3,5\}$ iterations of Algorithm \ref{alg:tandem} for $\eta=10^{-2}$ in Algorithm \ref{alg:restricted}}. The channel rates are equal to three bits and the per channel SNR is $\mathcal{E}_b=-10$ dB. The system\revise{s are} initialized in such a way that each DM, regardless of its observation, passes its input from its predecessor to its successor, i.e., $\gamma_k(x_k, u_{k-1})=u_{k-1}$, \, $1< k <N$ or equivalently
\begin{equation*}
{\bf{P}}^{k}_j=\mathbf{I}_{\Vert 2^R \Vert}\,,
\end{equation*}
and the first DM provides its output randomly from the index set $\{1, \ldots, 2^R\}$. Then, in the initialized network, the fusion center (DM $N$) uses the MAP criterion to make the global decision, which gives the same error probability regardless of the number of DMs before it. The proposed algorithm results in a significant performance improvement after the first iteration and shows no visible improvement after three iterations \revise{for $\eta=10^{-6}$, and after five iterations for $\eta=10^{-2}$} over the range of $N=1,\ldots,20$. \revise{As is depicted in Fig.~\ref{fig:conv}, by relaxing the parameter $\eta$ in the inner algorithm (Algorithm \ref{alg:restricted}), the outer algorithm (Algorithm \ref{alg:tandem}) needs to perform more iterations to give the same performance.}
\begin{figure}[t]
\centering
\input{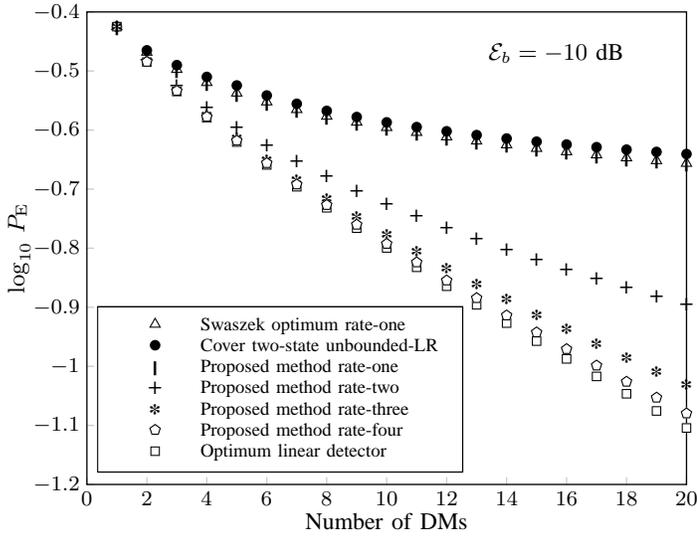}
\caption{Comparison of error probability performance of tandem networks with different channel rates and for different number of DMs after three iterations of design with unconstrained tandem network and existing methods for rate-one channels, for binary hypothesis testing problem and $\mathcal{E}_b=-10$ dB.}
\label{fig:perf}
\end{figure}

In Fig.~\ref{fig:perf} and Fig.~\ref{fig:perf1} the performance of the designed tandem network for various channel rates and number of DMs is compared to the optimum rate-one performance \cite{Swa93} and Cover's \cite{Cov69} rate-one method, for $\mathcal{E}_b=-10$ dB and $\mathcal{E}_b=0$ dB. For rate-one channels and under a few symmetry conditions satisfied in our simulation setup, Swaszek \cite{Swa93} found that the optimum distributed tests, satisfying the set of necessary conditions presented in \cite{Ek82,Varsh96}, coincide with the optimal local tests and are given by
\begin{equation*}
x_i\overset{u_i=1}{\underset{u_i=0}{\gtrless}}(1-2u_{i-1})\tau_i\,,
\end{equation*}
where $u_{i-1}\in\{0,1\}$ and where the threshold $\tau_i$ is found from the log-likelihood ratio test at each DM as
\begin{equation}
\tau_i=\frac{\sigma^2}{2a}\ln\left\{\frac{1-P_\mathrm{E}(i-1)}{P_\mathrm{E}(i-1)} \right\}\,,
\label{eq:swatau}\end{equation}
where $P_\mathrm{E}(i)$ is the error probability at the output of DM $i$. Swaszek \cite{Swa93} also found a recursive expression for the minimum error probability with rate-one channels given by
\begin{equation}
\begin{split}
P_\mathrm{E}(i)=&\,\mathcal{Q}\!\left(\frac{a+\tau_i}{\sigma}\right)+\\
&P_\mathrm{E}(i-1)\left[ \mathcal{Q}\!\left(\frac{a-\tau_i}{\sigma}\right)-\mathcal{Q}\!\left(\frac{a+\tau_i}{\sigma}\right)\right]\,,
\label{eq:optPE}\end{split}\end{equation}
where $\mathcal{Q}(x)$ is the tail probability of unit-variance Gaussian density. To prove a point regarding vanishing error probabilities under unbounded likelihood \revise{ratios}, Cover \cite{Cov69} had previously proposed a test for resolving $-a$ versus $+a$ in Gaussian noise, given by
\begin{equation*}
  u_i=\begin{cases}
        1  & x_i>+\tau_i \\
        0  & x_i<-\tau_i \\
        u_{i-1} & \text{otherwise}\,,\\
  \end{cases}
\end{equation*}
where $u_0$ is chosen arbitrary from $\{0,1\}$ and where
\begin{equation}
\tau_i=\sqrt{2\sigma^2\log_{10}i}\,.
\label{eq:covtau}\end{equation}

Plugging $\tau_i$ from \eqref{eq:swatau} or from \eqref{eq:covtau} into \eqref{eq:optPE} provides a recursive method of calculating the probability of error of each method. These are shown for comparison in Fig.~\ref{fig:perf} and Fig.~\ref{fig:perf1}. We also include the performance of the unconstrained linear detector which is optimum for this problem when the channels are infinite-rate ($R=\infty$). The linear detector is optimal for the \revise{Gaussian} observation model and is given by \cite{Swa93}
\begin{equation}
\sum_{i=1}^{N} x_i\overset{u_N=1}{\underset{u_N=0}{\gtrless}}0\, . \label{eq:optimumlinear}
\end{equation}
\begin{figure}[t]
\centering
\psset{xunit=0.071429\plotwidth,yunit=0.225346\plotwidth}%
\begin{pspicture}(-1.677419,-4.388889)(14.258065,-0.418129)%


\psline[linewidth=\AxesLineWidth,linecolor=GridColor](0.000000,-4.000000)(0.000000,-3.946748)
\psline[linewidth=\AxesLineWidth,linecolor=GridColor](2.000000,-4.000000)(2.000000,-3.946748)
\psline[linewidth=\AxesLineWidth,linecolor=GridColor](4.000000,-4.000000)(4.000000,-3.946748)
\psline[linewidth=\AxesLineWidth,linecolor=GridColor](6.000000,-4.000000)(6.000000,-3.946748)
\psline[linewidth=\AxesLineWidth,linecolor=GridColor](8.000000,-4.000000)(8.000000,-3.946748)
\psline[linewidth=\AxesLineWidth,linecolor=GridColor](10.000000,-4.000000)(10.000000,-3.946748)
\psline[linewidth=\AxesLineWidth,linecolor=GridColor](12.000000,-4.000000)(12.000000,-3.946748)
\psline[linewidth=\AxesLineWidth,linecolor=GridColor](14.000000,-4.000000)(14.000000,-3.946748)
\psline[linewidth=\AxesLineWidth,linecolor=GridColor](0.000000,-4.000000)(0.168000,-4.000000)
\psline[linewidth=\AxesLineWidth,linecolor=GridColor](0.000000,-3.500000)(0.168000,-3.500000)
\psline[linewidth=\AxesLineWidth,linecolor=GridColor](0.000000,-3.000000)(0.168000,-3.000000)
\psline[linewidth=\AxesLineWidth,linecolor=GridColor](0.000000,-2.500000)(0.168000,-2.500000)
\psline[linewidth=\AxesLineWidth,linecolor=GridColor](0.000000,-2.000000)(0.168000,-2.000000)
\psline[linewidth=\AxesLineWidth,linecolor=GridColor](0.000000,-1.500000)(0.168000,-1.500000)
\psline[linewidth=\AxesLineWidth,linecolor=GridColor](0.000000,-1.000000)(0.168000,-1.000000)
\psline[linewidth=\AxesLineWidth,linecolor=GridColor](0.000000,-0.500000)(0.168000,-0.500000)

{ \footnotesize 
\rput[t](0.000000,-4.053252){$0$}
\rput[t](2.000000,-4.053252){$2$}
\rput[t](4.000000,-4.053252){$4$}
\rput[t](6.000000,-4.053252){$6$}
\rput[t](8.000000,-4.053252){$8$}
\rput[t](10.000000,-4.053252){$10$}
\rput[t](12.000000,-4.053252){$12$}
\rput[t](14.000000,-4.053252){$14$}
\rput[r](-0.168000,-4.000000){$-4$}
\rput[r](-0.168000,-3.500000){$-3.5$}
\rput[r](-0.168000,-3.000000){$-3$}
\rput[r](-0.168000,-2.500000){$-2.5$}
\rput[r](-0.168000,-2.000000){$-2$}
\rput[r](-0.168000,-1.500000){$-1.5$}
\rput[r](-0.168000,-1.000000){$-1$}
\rput[r](-0.168000,-0.500000){$-0.5$}
} 

\psframe[linewidth=\AxesLineWidth,dimen=middle](0.000000,-4.000000)(14.000000,-0.500000)

{ \small 
\rput[b](7.000000,-4.388889){
\begin{tabular}{c}
Number of DMs\\
\end{tabular}
}

\rput[t]{90}(-1.9,-2.250000){
\begin{tabular}{c}
$\log_{10}$ $P_\mathrm{E}$\\
\end{tabular}
}
} 

\newrgbcolor{color288.0039}{0  0  0}
\psline[plotstyle=line,linejoin=1,showpoints=false,dotstyle=Btriangle,dotsize=\MarkerSize,linestyle=none,linewidth=\LineWidth,linecolor=color288.0039]
(14.000000,-1.841077)(14.000000,-1.841077)
\psline[plotstyle=line,linejoin=1,showpoints=true,dotstyle=Btriangle,dotsize=\MarkerSize,linestyle=none,linewidth=\LineWidth,linecolor=color288.0039]
(1.000000,-0.799546)(2.000000,-1.013639)(3.000000,-1.160000)(4.000000,-1.281686)(5.000000,-1.376898)
(6.000000,-1.456945)(7.000000,-1.525814)(8.000000,-1.586110)(9.000000,-1.639628)(10.000000,-1.687660)
(11.000000,-1.731165)(12.000000,-1.770875)(13.000000,-1.807362)(14.000000,-1.841077)

\newrgbcolor{color289.0034}{0  0  0}
\psline[plotstyle=line,linejoin=1,showpoints=false,dotstyle=*,dotsize=\MarkerSize,linestyle=none,linewidth=\LineWidth,linecolor=color289.0034]
(14.000000,-1.684159)(14.000000,-1.684159)
\psline[plotstyle=line,linejoin=1,showpoints=true,dotstyle=*,dotsize=\MarkerSize,linestyle=none,linewidth=\LineWidth,linecolor=color289.0034]
(1.000000,-0.799546)(2.000000,-1.008422)(3.000000,-1.160770)(4.000000,-1.275199)(5.000000,-1.362816)
(6.000000,-1.431133)(7.000000,-1.485432)(8.000000,-1.529455)(9.000000,-1.565862)(10.000000,-1.596546)
(11.000000,-1.622863)(12.000000,-1.645791)(13.000000,-1.666047)(14.000000,-1.684159)

\newrgbcolor{color290.0034}{0  0  0}
\psline[plotstyle=line,linejoin=1,showpoints=false,dotstyle=Bsquare,dotsize=\MarkerSize,linestyle=none,linewidth=\LineWidth,linecolor=color290.0034]
(13.000000,-3.807585)(13.831369,-4.000000)
\psline[plotstyle=line,linejoin=1,showpoints=true,dotstyle=Bsquare,dotsize=\MarkerSize,linestyle=none,linewidth=\LineWidth,linecolor=color290.0034]
(1.000000,-0.799546)(2.000000,-1.104303)(3.000000,-1.380570)(4.000000,-1.643016)(5.000000,-1.897098)
(6.000000,-2.145515)(7.000000,-2.389821)(8.000000,-2.630994)(9.000000,-2.869699)(10.000000,-3.106404)
(11.000000,-3.341455)(12.000000,-3.575114)(13.000000,-3.807585)

\newrgbcolor{color291.0034}{0  0  0}
\psline[plotstyle=line,linejoin=1,showpoints=false,dotstyle=B|,dotsize=\MarkerSize,linestyle=none,linewidth=\LineWidth,linecolor=color291.0034]
(14.000000,-1.827470)(14.000000,-1.827470)
\psline[plotstyle=line,linejoin=1,showpoints=true,dotstyle=B|,dotsize=\MarkerSize,linestyle=none,linewidth=\LineWidth,linecolor=color291.0034]
(1.000000,-0.799546)(2.000000,-1.013184)(3.000000,-1.163865)(4.000000,-1.280339)(5.000000,-1.374789)(6.000000,-1.454190)
(7.000000,-1.522205)(8.000000,-1.581481)(9.000000,-1.633878)(10.000000,-1.680677)(11.000000,-1.722751)
(12.000000,-1.760837)(13.000000,-1.795675)(14.000000,-1.827470)

\newrgbcolor{color292.0034}{0  0  0}
\psline[plotstyle=line,linejoin=1,showpoints=false,dotstyle=B+,dotsize=\MarkerSize,linestyle=none,linewidth=\LineWidth,linecolor=color292.0034]
(14.000000,-2.849704)(14.000000,-2.849704)
\psline[plotstyle=line,linejoin=1,showpoints=true,dotstyle=B+,dotsize=\MarkerSize,linestyle=none,linewidth=\LineWidth,linecolor=color292.0034]
(1.000000,-0.799546)(2.000000,-1.079948)(3.000000,-1.319933)(4.000000,-1.534977)(5.000000,-1.731480)(6.000000,-1.911663)
(7.000000,-2.076329)(8.000000,-2.227892)(9.000000,-2.365542)(10.000000,-2.489611)(11.000000,-2.600043)
(12.000000,-2.696965)(13.000000,-2.779892)(14.000000,-2.849704)

\newrgbcolor{color293.0034}{0  0  0}
\psline[plotstyle=line,linejoin=1,showpoints=false,dotstyle=Basterisk,dotsize=\MarkerSize,linestyle=none,linewidth=\LineWidth,linecolor=color293.0034]
(14.000000,-3.194749)(14.000000,-3.194749)
\psline[plotstyle=line,linejoin=1,showpoints=true,dotstyle=Basterisk,dotsize=\MarkerSize,linestyle=none,linewidth=\LineWidth,linecolor=color293.0034]
(1.000000,-0.799546)(2.000000,-1.097596)(3.000000,-1.363352)(4.000000,-1.612662)(5.000000,-1.848881)(6.000000,-2.073577)
(7.000000,-2.287020)(8.000000,-2.487186)(9.000000,-2.670071)(10.000000,-2.831778)(11.000000,-2.967151)
(12.000000,-3.072799)(13.000000,-3.147917)(14.000000,-3.194749)

\newrgbcolor{color294.0034}{0  0  0}
\psline[plotstyle=line,linejoin=1,showpoints=false,dotstyle=Bpentagon,dotsize=\MarkerSize,linestyle=none,linewidth=\LineWidth,linecolor=color294.0034]
(14.000000,-3.253377)(14.000000,-3.253377)
\psline[plotstyle=line,linejoin=1,showpoints=true,dotstyle=Bpentagon,dotsize=\MarkerSize,linestyle=none,linewidth=\LineWidth,linecolor=color294.0034]
(1.000000,-0.799546)(2.000000,-1.101597)(3.000000,-1.374439)(4.000000,-1.631631)(5.000000,-1.878464)(6.000000,-2.116764)
(7.000000,-2.344777)(8.000000,-2.559048)(9.000000,-2.755505)(10.000000,-2.925307)(11.000000,-3.061394)
(12.000000,-3.159744)(13.000000,-3.221749)(14.000000,-3.253377)

{ \scriptsize  
\rput(4.45,-3.3009430){%
\psshadowbox[framesep=0pt,shadowsize=0pt,linewidth=\AxesLineWidth]{\psframebox*{\begin{tabular}{l}
\Rnode{a1}{\hspace*{0.0ex}} \hspace*{0.7cm} \Rnode{a2}{~~Swaszek optimum rate-one} \\
\Rnode{a3}{\hspace*{0.0ex}} \hspace*{0.7cm} \Rnode{a4}{~~Cover two-state unbounded-LR} \\
\Rnode{a7}{\hspace*{0.0ex}} \hspace*{0.7cm} \Rnode{a8}{~~Proposed method rate-one} \\
\Rnode{a9}{\hspace*{0.0ex}} \hspace*{0.7cm} \Rnode{a10}{~~Proposed method rate-two} \\
\Rnode{a11}{\hspace*{0.0ex}} \hspace*{0.7cm} \Rnode{a12}{~~Proposed method rate-three} \\
\Rnode{a13}{\hspace*{0.0ex}} \hspace*{0.7cm} \Rnode{a14}{~~Proposed method rate-four} \\
\Rnode{a5}{\hspace*{0.0ex}} \hspace*{0.7cm} \Rnode{a6}{~~Optimum linear detector} \\
\end{tabular}}
\ncline[linestyle=none,linewidth=\LineWidth,linecolor=color288.0039]{a1}{a2} \ncput{\psdot[dotstyle=Btriangle,dotsize=\MarkerSize,linecolor=color288.0039]}
\ncline[linestyle=none,linewidth=\LineWidth,linecolor=color289.0034]{a3}{a4} \ncput{\psdot[dotstyle=*,dotsize=\MarkerSize,linecolor=color289.0034]}
\ncline[linestyle=none,linewidth=\LineWidth,linecolor=color290.0034]{a5}{a6} \ncput{\psdot[dotstyle=Bsquare,dotsize=\MarkerSize,linecolor=color290.0034]}
\ncline[linestyle=none,linewidth=\LineWidth,linecolor=color291.0034]{a7}{a8} \ncput{\psdot[dotstyle=B|,dotsize=\MarkerSize,linecolor=color291.0034]}
\ncline[linestyle=none,linewidth=\LineWidth,linecolor=color292.0034]{a9}{a10} \ncput{\psdot[dotstyle=B+,dotsize=\MarkerSize,linecolor=color292.0034]}
\ncline[linestyle=none,linewidth=\LineWidth,linecolor=color293.0034]{a11}{a12} \ncput{\psdot[dotstyle=Basterisk,dotsize=\MarkerSize,linecolor=color293.0034]}
\ncline[linestyle=none,linewidth=\LineWidth,linecolor=color294.0034]{a13}{a14} \ncput{\psdot[dotstyle=Bpentagon,dotsize=\MarkerSize,linecolor=color294.0034]}
}%
}%
} 

{ \small 
\newrgbcolor{color220.0034}{0  0  0}
\uput{0pt}[0](9.517103,-.80630){%
\psframebox[framesep=1pt,fillstyle=solid,linestyle=none,linewidth=0.5pt]{\begin{tabular}{@{}c@{}}
$\mathcal{E}_b= 0$ dB\\[-0.3ex]
\end{tabular}}}
} 

\end{pspicture}%
\caption{Comparison of error probability performance of tandem networks with different channel rates and for different number of DMs after three iterations of design with unconstrained tandem network and existing methods for rate-one channels, for binary hypothesis testing problem and $\mathcal{E}_b=0$ dB.}
\label{fig:perf1}
\end{figure}
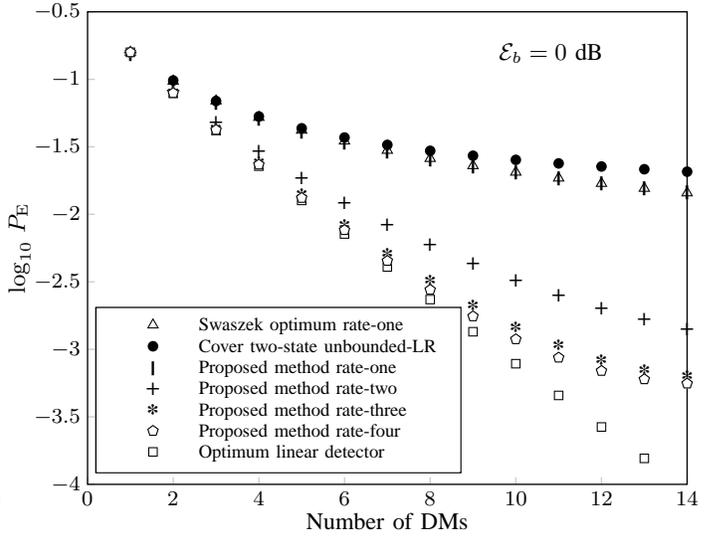
\begin{figure}[b]
\centering
%

\psset{xunit=0.100000\plotwidth,yunit=0.394355\plotwidth}%
\begin{pspicture}(-11.198157,-2.722222)(0.115207,-0.453216)%


\psline[linewidth=\AxesLineWidth,linecolor=GridColor](-10.000000,-2.500000)(-10.000000,-2.469571)
\psline[linewidth=\AxesLineWidth,linecolor=GridColor](-9.000000,-2.500000)(-9.000000,-2.469571)
\psline[linewidth=\AxesLineWidth,linecolor=GridColor](-8.000000,-2.500000)(-8.000000,-2.469571)
\psline[linewidth=\AxesLineWidth,linecolor=GridColor](-7.000000,-2.500000)(-7.000000,-2.469571)
\psline[linewidth=\AxesLineWidth,linecolor=GridColor](-6.000000,-2.500000)(-6.000000,-2.469571)
\psline[linewidth=\AxesLineWidth,linecolor=GridColor](-5.000000,-2.500000)(-5.000000,-2.469571)
\psline[linewidth=\AxesLineWidth,linecolor=GridColor](-4.000000,-2.500000)(-4.000000,-2.469571)
\psline[linewidth=\AxesLineWidth,linecolor=GridColor](-3.000000,-2.500000)(-3.000000,-2.469571)
\psline[linewidth=\AxesLineWidth,linecolor=GridColor](-2.000000,-2.500000)(-2.000000,-2.469571)
\psline[linewidth=\AxesLineWidth,linecolor=GridColor](-1.000000,-2.500000)(-1.000000,-2.469571)
\psline[linewidth=\AxesLineWidth,linecolor=GridColor](0.000000,-2.500000)(0.000000,-2.469571)
\psline[linewidth=\AxesLineWidth,linecolor=GridColor](-10.000000,-2.500000)(-9.880000,-2.500000)
\psline[linewidth=\AxesLineWidth,linecolor=GridColor](-10.000000,-2.000000)(-9.880000,-2.000000)
\psline[linewidth=\AxesLineWidth,linecolor=GridColor](-10.000000,-1.500000)(-9.880000,-1.500000)
\psline[linewidth=\AxesLineWidth,linecolor=GridColor](-10.000000,-1.000000)(-9.880000,-1.000000)
\psline[linewidth=\AxesLineWidth,linecolor=GridColor](-10.000000,-0.500000)(-9.880000,-0.500000)

{ \footnotesize 
\rput[t](-10.000000,-2.530429){$-10$}
\rput[t](-9.000000,-2.530429){$-9$}
\rput[t](-8.000000,-2.530429){$-8$}
\rput[t](-7.000000,-2.530429){$-7$}
\rput[t](-6.000000,-2.530429){$-6$}
\rput[t](-5.000000,-2.530429){$-5$}
\rput[t](-4.000000,-2.530429){$-4$}
\rput[t](-3.000000,-2.530429){$-3$}
\rput[t](-2.000000,-2.530429){$-2$}
\rput[t](-1.000000,-2.530429){$-1$}
\rput[t](0.000000,-2.530429){$0$}
\rput[r](-10.120000,-2.500000){$-2.5$}
\rput[r](-10.120000,-2.000000){$-2$}
\rput[r](-10.120000,-1.500000){$-1.5$}
\rput[r](-10.120000,-1.000000){$-1$}
\rput[r](-10.120000,-0.500000){$-0.5$}
} 

\psframe[linewidth=\AxesLineWidth,dimen=middle](-10.000000,-2.500000)(0.000000,-0.500000)

{ \small 
\rput[b](-5.000000,-2.722222){
\begin{tabular}{c}
$\mathcal{E}_b$ (dB)\\
\end{tabular}
}

\rput[t]{90}(-11.4,-1.500000){
\begin{tabular}{c}
$\log_{10}$ $P_\mathrm{E}$\\
\end{tabular}
}
} 

\newrgbcolor{color267.0016}{0  0  0}
\psline[plotstyle=line,linejoin=1,showpoints=false,dotstyle=Bpentagon,dotsize=\MarkerSize,linestyle=solid,linewidth=\LineWidth,linecolor=color267.0016]
(0.000000,-1.522205)(0.000000,-1.522205)
\psline[plotstyle=line,linejoin=1,showpoints=true,dotstyle=Bpentagon,dotsize=\MarkerSize,linestyle=solid,linewidth=\LineWidth,linecolor=color267.0016]
(-10.000000,-0.564862)(-9.000000,-0.604679)(-8.000000,-0.651398)(-7.000000,-0.706349)(-6.000000,-0.771302)
(-5.000000,-0.848248)(-4.000000,-0.939729)(-3.000000,-1.048763)(-2.000000,-1.179011)(-1.000000,-1.335049)
(0.000000,-1.522205)

\newrgbcolor{color268.0011}{0  0  0}
\psline[plotstyle=line,linejoin=1,showpoints=false,dotstyle=B+,dotsize=\MarkerSize,linestyle=solid,linewidth=\LineWidth,linecolor=color268.0011]
(0.000000,-2.075721)(0.000000,-2.075721)
\psline[plotstyle=line,linejoin=1,showpoints=true,dotstyle=B+,dotsize=\MarkerSize,linestyle=solid,linewidth=\LineWidth,linecolor=color268.0011]
(-10.000000,-0.652947)(-9.000000,-0.708559)(-8.000000,-0.774481)(-7.000000,-0.852696)(-6.000000,-0.946153)
(-5.000000,-1.057992)(-4.000000,-1.192465)(-3.000000,-1.355561)(-2.000000,-1.551294)(-1.000000,-1.787812)
(0.000000,-2.075721)

\newrgbcolor{color269.0011}{0  0  0}
\psline[plotstyle=line,linejoin=1,showpoints=false,dotstyle=B|,dotsize=\MarkerSize,linestyle=solid,linewidth=\LineWidth,linecolor=color269.0011]
(0.000000,-2.286073)(0.000000,-2.286073)
\psline[plotstyle=line,linejoin=1,showpoints=true,dotstyle=B|,dotsize=\MarkerSize,linestyle=solid,linewidth=\LineWidth,linecolor=color269.0011]
(-10.000000,-0.683463)(-9.000000,-0.744735)(-8.000000,-0.817601)(-7.000000,-0.904565)(-6.000000,-1.008551)
(-5.000000,-1.133819)(-4.000000,-1.284732)(-3.000000,-1.467513)(-2.000000,-1.689774)(-1.000000,-1.959437)
(0.000000,-2.286073)

\newrgbcolor{color270.0011}{0  0  0}
\psline[plotstyle=line,linejoin=1,showpoints=false,dotstyle=*,dotsize=\MarkerSize,linestyle=solid,linewidth=\LineWidth,linecolor=color270.0011]
(0.000000,-2.344074)(0.000000,-2.344074)
\psline[plotstyle=line,linejoin=1,showpoints=true,dotstyle=*,dotsize=\MarkerSize,linestyle=solid,linewidth=\LineWidth,linecolor=color270.0011]
(-10.000000,-0.692026)(-9.000000,-0.755056)(-8.000000,-0.829803)(-7.000000,-0.919166)(-6.000000,-1.026324)
(-5.000000,-1.155169)(-4.000000,-1.310904)(-3.000000,-1.499530)(-2.000000,-1.729158)(-1.000000,-2.008034)
(0.000000,-2.344074)

\newrgbcolor{color271.0011}{0  0  0}
\psline[plotstyle=line,linejoin=1,linestyle=solid,linewidth=\LineWidth,linecolor=color271.0011]
(0.000000,-2.389821)(0.000000,-2.389821)
\psline[plotstyle=line,linejoin=1,linestyle=solid,linewidth=\LineWidth,linecolor=color271.0011]
(-10.000000,-0.695958)(-9.000000,-0.759625)(-8.000000,-0.835339)(-7.000000,-0.925770)(-6.000000,-1.034248)
(-5.000000,-1.164945)(-4.000000,-1.323092)(-3.000000,-1.515261)(-2.000000,-1.749717)(-1.000000,-2.036867)
(0.000000,-2.389821)

{ \scriptsize 
\rput(-8,-2.15){%
\psshadowbox[framesep=0pt,shadowsize=0pt,linewidth=\AxesLineWidth]{\psframebox*{\begin{tabular}{l}
\Rnode{a1}{\hspace*{0.0ex}} \hspace*{0.7cm} \Rnode{a2}{~~$R$ = $1$ bit} \\
\Rnode{a3}{\hspace*{0.0ex}} \hspace*{0.7cm} \Rnode{a4}{~~$R$ = $2$ bits} \\
\Rnode{a5}{\hspace*{0.0ex}} \hspace*{0.7cm} \Rnode{a6}{~~$R$ = $3$ bits} \\
\Rnode{a7}{\hspace*{0.0ex}} \hspace*{0.7cm} \Rnode{a8}{~~$R$ = $4$ bits} \\
\Rnode{a9}{\hspace*{0.0ex}} \hspace*{0.7cm} \Rnode{a10}{~~$R$ = $\infty$} \\
\end{tabular}}
\ncline[linestyle=solid,linewidth=\LineWidth,linecolor=color267.0016]{a1}{a2} \ncput{\psdot[dotstyle=Bpentagon,dotsize=\MarkerSize,linecolor=color267.0016]}
\ncline[linestyle=solid,linewidth=\LineWidth,linecolor=color268.0011]{a3}{a4} \ncput{\psdot[dotstyle=B+,dotsize=\MarkerSize,linecolor=color268.0011]}
\ncline[linestyle=solid,linewidth=\LineWidth,linecolor=color269.0011]{a5}{a6} \ncput{\psdot[dotstyle=B|,dotsize=\MarkerSize,linecolor=color269.0011]}
\ncline[linestyle=solid,linewidth=\LineWidth,linecolor=color270.0011]{a7}{a8} \ncput{\psdot[dotstyle=*,dotsize=\MarkerSize,linecolor=color270.0011]}
\ncline[linestyle=solid,linewidth=\LineWidth,linecolor=color271.0011]{a9}{a10}
}%
}%
} 

\end{pspicture}%
\caption{\revise{Performance of designed tandem network with $N=7$ DMs as a function of channels SNR $\mathcal{E}_b$.}}
\label{fig:snr}
\end{figure}
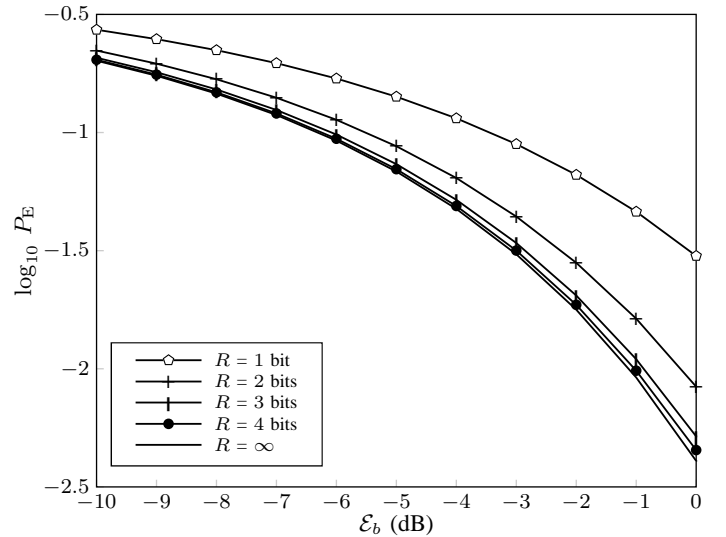
Since \eqref{eq:optimumlinear} can be recursively computed over the DMs in the absence of rate constraints the optimum centralized and decentralized solutions coincide. The error probability of the unconstrained serial network with $N$ DMs for binary hypothesis testing problem is accordingly equal to $P_\mathrm{E}=\mathcal{Q}(a\sqrt{N}/\sigma)$.

The results of the proposed method in Fig.~\ref{fig:perf} and Fig.~\ref{fig:perf1} are achieved after $K=3$ iterations. For rate-one channels, the performance of the proposed method is indistinguishable from the optimum solution \cite{Swa93}, while increasing the channels rate leads to better performance which is in harmony with the parallel network. The simulation results show that increasing the rate of the channels between the DMs can significantly improve the performance of the tandem network: for example when the channel rates are equal to $R=4$ bits the performance of the designed network is very close to the unconstrained case, at least for $N$ up to $20$ for $\mathcal{E}_b=-10$ dB and for $N$ up to $10$ for $\mathcal{E}_b=0$ dB.

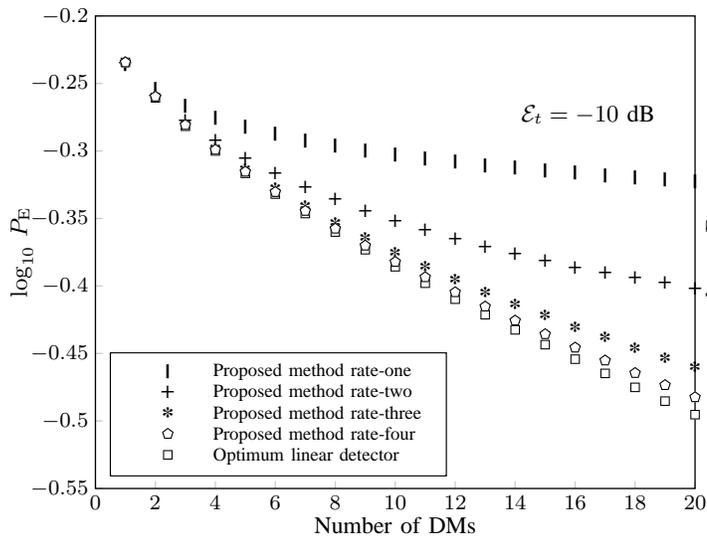
\begin{figure}[t]
\centering
%
%
\psset{xunit=0.050000\plotwidth,yunit=2.253456\plotwidth}%
\begin{pspicture}(-2.718894,-0.588889)(20.368664,-0.191813)%


\psline[linewidth=\AxesLineWidth,linecolor=GridColor](0.000000,-0.550000)(0.000000,-0.544675)
\psline[linewidth=\AxesLineWidth,linecolor=GridColor](2.000000,-0.550000)(2.000000,-0.544675)
\psline[linewidth=\AxesLineWidth,linecolor=GridColor](4.000000,-0.550000)(4.000000,-0.544675)
\psline[linewidth=\AxesLineWidth,linecolor=GridColor](6.000000,-0.550000)(6.000000,-0.544675)
\psline[linewidth=\AxesLineWidth,linecolor=GridColor](8.000000,-0.550000)(8.000000,-0.544675)
\psline[linewidth=\AxesLineWidth,linecolor=GridColor](10.000000,-0.550000)(10.000000,-0.544675)
\psline[linewidth=\AxesLineWidth,linecolor=GridColor](12.000000,-0.550000)(12.000000,-0.544675)
\psline[linewidth=\AxesLineWidth,linecolor=GridColor](14.000000,-0.550000)(14.000000,-0.544675)
\psline[linewidth=\AxesLineWidth,linecolor=GridColor](16.000000,-0.550000)(16.000000,-0.544675)
\psline[linewidth=\AxesLineWidth,linecolor=GridColor](18.000000,-0.550000)(18.000000,-0.544675)
\psline[linewidth=\AxesLineWidth,linecolor=GridColor](20.000000,-0.550000)(20.000000,-0.544675)
\psline[linewidth=\AxesLineWidth,linecolor=GridColor](0.000000,-0.550000)(0.240000,-0.550000)
\psline[linewidth=\AxesLineWidth,linecolor=GridColor](0.000000,-0.500000)(0.240000,-0.500000)
\psline[linewidth=\AxesLineWidth,linecolor=GridColor](0.000000,-0.450000)(0.240000,-0.450000)
\psline[linewidth=\AxesLineWidth,linecolor=GridColor](0.000000,-0.400000)(0.240000,-0.400000)
\psline[linewidth=\AxesLineWidth,linecolor=GridColor](0.000000,-0.350000)(0.240000,-0.350000)
\psline[linewidth=\AxesLineWidth,linecolor=GridColor](0.000000,-0.300000)(0.240000,-0.300000)
\psline[linewidth=\AxesLineWidth,linecolor=GridColor](0.000000,-0.250000)(0.240000,-0.250000)
\psline[linewidth=\AxesLineWidth,linecolor=GridColor](0.000000,-0.200000)(0.240000,-0.200000)

{ \footnotesize 
\rput[t](0.000000,-0.555325){$0$}
\rput[t](2.000000,-0.555325){$2$}
\rput[t](4.000000,-0.555325){$4$}
\rput[t](6.000000,-0.555325){$6$}
\rput[t](8.000000,-0.555325){$8$}
\rput[t](10.000000,-0.555325){$10$}
\rput[t](12.000000,-0.555325){$12$}
\rput[t](14.000000,-0.555325){$14$}
\rput[t](16.000000,-0.555325){$16$}
\rput[t](18.000000,-0.555325){$18$}
\rput[t](20.000000,-0.555325){$20$}
\rput[r](-0.240000,-0.550000){$-0.55$}
\rput[r](-0.240000,-0.500000){$-0.5$}
\rput[r](-0.240000,-0.450000){$-0.45$}
\rput[r](-0.240000,-0.400000){$-0.4$}
\rput[r](-0.240000,-0.350000){$-0.35$}
\rput[r](-0.240000,-0.300000){$-0.3$}
\rput[r](-0.240000,-0.250000){$-0.25$}
\rput[r](-0.240000,-0.200000){$-0.2$}
} 

\psframe[linewidth=\AxesLineWidth,dimen=middle](0.000000,-0.550000)(20.000000,-0.200000)

{ \small 
\rput[b](10.000000,-0.588889){
\begin{tabular}{c}
Number of DMs\\
\end{tabular}
}

\rput[t]{90}(-2.9,-0.375000){
\begin{tabular}{c}
$\log_{10}$ $P_\mathrm{E}$\\
\end{tabular}
}
} 

\newrgbcolor{color335.0016}{0  0  0}
\psline[plotstyle=line,linejoin=1,showpoints=true,dotstyle=B|,dotsize=\MarkerSize,linestyle=none,linewidth=\LineWidth,linecolor=color335.0016]
(1.000000,-0.234397)(2.000000,-0.253210)(3.000000,-0.265200)(4.000000,-0.274007)(5.000000,-0.280835)(6.000000,-0.286425)
(7.000000,-0.291069)(8.000000,-0.295078)(9.000000,-0.298605)(10.000000,-0.301725)(11.000000,-0.304518)
(12.000000,-0.306977)(13.000000,-0.309272)(14.000000,-0.311402)(15.000000,-0.313274)(16.000000,-0.315065)
(17.000000,-0.316773)(18.000000,-0.318307)(19.000000,-0.319755)(20.000000,-0.321118)

\newrgbcolor{color336.0011}{0  0  0}
\psline[plotstyle=line,linejoin=1,showpoints=true,dotstyle=B+,dotsize=\MarkerSize,linestyle=none,linewidth=\LineWidth,linecolor=color336.0011]
(1.000000,-0.234397)(2.000000,-0.258612)(3.000000,-0.277037)(4.000000,-0.292260)(5.000000,-0.305307)(6.000000,-0.316683)
(7.000000,-0.326795)(8.000000,-0.335828)(9.000000,-0.344094)(10.000000,-0.351640)(11.000000,-0.358526)
(12.000000,-0.364617)(13.000000,-0.370692)(14.000000,-0.376234)(15.000000,-0.381011)(16.000000,-0.386053)
(17.000000,-0.390192)(18.000000,-0.393834)(19.000000,-0.397397)(20.000000,-0.402086)

\newrgbcolor{color337.0011}{0  0  0}
\psline[plotstyle=line,linejoin=1,showpoints=true,dotstyle=Bsquare,dotsize=\MarkerSize,linestyle=none,linewidth=\LineWidth,linecolor=color337.0011]
(1.000000,-0.234397)(2.000000,-0.260658)(3.000000,-0.281669)(4.000000,-0.299972)(5.000000,-0.316549)
(6.000000,-0.331903)(7.000000,-0.346334)(8.000000,-0.360035)(9.000000,-0.373142)(10.000000,-0.385753)
(11.000000,-0.397942)(12.000000,-0.409766)(13.000000,-0.421270)(14.000000,-0.432493)(15.000000,-0.443463)
(16.000000,-0.454207)(17.000000,-0.464746)(18.000000,-0.475098)(19.000000,-0.485279)(20.000000,-0.495303)

\newrgbcolor{color338.0011}{0  0  0}
\psline[plotstyle=line,linejoin=1,showpoints=true,dotstyle=Basterisk,dotsize=\MarkerSize,linestyle=none,linewidth=\LineWidth,linecolor=color338.0011]
(1.000000,-0.234397)(2.000000,-0.259716)(3.000000,-0.280337)(4.000000,-0.297742)(5.000000,-0.313274)(6.000000,-0.327348)
(7.000000,-0.340464)(8.000000,-0.352715)(9.000000,-0.364114)(10.000000,-0.374996)(11.000000,-0.385314)
(12.000000,-0.395018)(13.000000,-0.404394)(14.000000,-0.413300)(15.000000,-0.421705)(16.000000,-0.429924)
(17.000000,-0.437826)(18.000000,-0.445390)(19.000000,-0.452841)(20.000000,-0.459921)

\newrgbcolor{color339.0011}{0  0  0}
\psline[plotstyle=line,linejoin=1,showpoints=true,dotstyle=Bpentagon,dotsize=\MarkerSize,linestyle=none,linewidth=\LineWidth,linecolor=color339.0011]
(1.000000,-0.234397)(2.000000,-0.259716)(3.000000,-0.280752)(4.000000,-0.299123)(5.000000,-0.315334)(6.000000,-0.330404)
(7.000000,-0.344381)(8.000000,-0.357535)(9.000000,-0.370081)(10.000000,-0.382161)(11.000000,-0.393511)
(12.000000,-0.404614)(13.000000,-0.415217)(14.000000,-0.425621)(15.000000,-0.435689)(16.000000,-0.445632)
(17.000000,-0.455188)(18.000000,-0.464453)(19.000000,-0.473402)(20.000000,-0.482408)

{ \scriptsize  
\rput(6.1,-0.495){%
\psshadowbox[framesep=0pt,shadowsize=0pt,linewidth=\AxesLineWidth]{\psframebox*{\begin{tabular}{l}
\Rnode{a1}{\hspace*{0.0ex}} \hspace*{0.7cm} \Rnode{a2}{~~Proposed method rate-one} \\
\Rnode{a3}{\hspace*{0.0ex}} \hspace*{0.7cm} \Rnode{a4}{~~Proposed method rate-two} \\
\Rnode{a7}{\hspace*{0.0ex}} \hspace*{0.7cm} \Rnode{a8}{~~Proposed method rate-three} \\
\Rnode{a9}{\hspace*{0.0ex}} \hspace*{0.7cm} \Rnode{a10}{~~Proposed method rate-four} \\
\Rnode{a5}{\hspace*{0.0ex}} \hspace*{0.7cm} \Rnode{a6}{~~Optimum linear detector} \\
\end{tabular}}
\ncline[linestyle=none,linewidth=\LineWidth,linecolor=color335.0016]{a1}{a2} \ncput{\psdot[dotstyle=B|,dotsize=\MarkerSize,linecolor=color335.0016]}
\ncline[linestyle=none,linewidth=\LineWidth,linecolor=color336.0011]{a3}{a4} \ncput{\psdot[dotstyle=B+,dotsize=\MarkerSize,linecolor=color336.0011]}
\ncline[linestyle=none,linewidth=\LineWidth,linecolor=color337.0011]{a5}{a6} \ncput{\psdot[dotstyle=Bsquare,dotsize=\MarkerSize,linecolor=color337.0011]}
\ncline[linestyle=none,linewidth=\LineWidth,linecolor=color338.0011]{a7}{a8} \ncput{\psdot[dotstyle=Basterisk,dotsize=\MarkerSize,linecolor=color338.0011]}
\ncline[linestyle=none,linewidth=\LineWidth,linecolor=color339.0011]{a9}{a10} \ncput{\psdot[dotstyle=Bpentagon,dotsize=\MarkerSize,linecolor=color339.0011]}
}%
}%
} 

{ \small 
\newrgbcolor{color208.0011}{0  0  0}
\uput{0pt}[0](14.067791,-0.272715){%
\psframebox[framesep=1pt,fillstyle=solid,linestyle=none,linewidth=0.5pt]{\begin{tabular}{@{}c@{}}
$\mathcal{E}_t = -10$ dB\\[-0.3ex]
\end{tabular}}}
} 

\end{pspicture}%
%
\caption{Comparison of error probability performance of tandem networks with different channel rates and for different number of DMs after three iterations of design with unconstrained tandem network, for ternary hypothesis testing problem and $\mathcal{E}_t=-10$ dB.}
\label{fig:M3}
\end{figure}

\revise{We further studied the performance of tandem network with $N=7$ DMs for different channels SNRs and rates. It is shown in Fig.~\ref{fig:snr} that increasing channels rate from $R=1$ to $R=2$ considerably improves the performance of the tandem network for the entire range $-10\text{ dB}\leq \mathcal{E}_b\leq 0\text{ dB}$, and the effect of increasing the channels rate at higher SNRs is more pronounced. This further motivates the necessity of a general method for the design of tandem network with multi-bit channels.}
\subsection{$M$-ary hypothesis testing} Next, we consider an $M$-ary hypothesis testing problem in which each real valued observation consists of a known signal $s_i$ in unit-variance additive white Gaussian Noise. We assume an equal distance signal set $\{s_1, \ldots,s_M\}$ in the interval $[-a,a]$, e.g., the test signal set is $\{-a, 0, a\}$ for ternary hypothesis testing and is $\{-a,-\frac{a}{3},\frac{a}{3},a\}$ for quaternary hypothesis testing. The observation model at each DM is
\begin{equation*}
H_i: {x_i}={s_i} +{n_i}, \quad i=1,\ldots,M\,.
\end{equation*}
We also define $\mathcal{E}_t=\vert a \vert^2$ and $\mathcal{E}_q=\vert a \vert^2$ for ternary and quaternary hypothesis test, respectively. As in the binary case, we assume that hypotheses are equally probable, i.e., $\pi_i=1/M$, $i=1, \ldots, M$, and that the channels between the DMs have the same rate.

\begin{figure}[t]
\centering
%
%
\psset{xunit=0.050000\plotwidth,yunit=4.381720\plotwidth}%
\begin{pspicture}(-2.718894,-0.360000)(20.368664,-0.155789)%


\psline[linewidth=\AxesLineWidth,linecolor=GridColor](0.000000,-0.340000)(0.000000,-0.337261)
\psline[linewidth=\AxesLineWidth,linecolor=GridColor](2.000000,-0.340000)(2.000000,-0.337261)
\psline[linewidth=\AxesLineWidth,linecolor=GridColor](4.000000,-0.340000)(4.000000,-0.337261)
\psline[linewidth=\AxesLineWidth,linecolor=GridColor](6.000000,-0.340000)(6.000000,-0.337261)
\psline[linewidth=\AxesLineWidth,linecolor=GridColor](8.000000,-0.340000)(8.000000,-0.337261)
\psline[linewidth=\AxesLineWidth,linecolor=GridColor](10.000000,-0.340000)(10.000000,-0.337261)
\psline[linewidth=\AxesLineWidth,linecolor=GridColor](12.000000,-0.340000)(12.000000,-0.337261)
\psline[linewidth=\AxesLineWidth,linecolor=GridColor](14.000000,-0.340000)(14.000000,-0.337261)
\psline[linewidth=\AxesLineWidth,linecolor=GridColor](16.000000,-0.340000)(16.000000,-0.337261)
\psline[linewidth=\AxesLineWidth,linecolor=GridColor](18.000000,-0.340000)(18.000000,-0.337261)
\psline[linewidth=\AxesLineWidth,linecolor=GridColor](20.000000,-0.340000)(20.000000,-0.337261)
\psline[linewidth=\AxesLineWidth,linecolor=GridColor](0.000000,-0.340000)(0.240000,-0.340000)
\psline[linewidth=\AxesLineWidth,linecolor=GridColor](0.000000,-0.320000)(0.240000,-0.320000)
\psline[linewidth=\AxesLineWidth,linecolor=GridColor](0.000000,-0.300000)(0.240000,-0.300000)
\psline[linewidth=\AxesLineWidth,linecolor=GridColor](0.000000,-0.280000)(0.240000,-0.280000)
\psline[linewidth=\AxesLineWidth,linecolor=GridColor](0.000000,-0.260000)(0.240000,-0.260000)
\psline[linewidth=\AxesLineWidth,linecolor=GridColor](0.000000,-0.240000)(0.240000,-0.240000)
\psline[linewidth=\AxesLineWidth,linecolor=GridColor](0.000000,-0.220000)(0.240000,-0.220000)
\psline[linewidth=\AxesLineWidth,linecolor=GridColor](0.000000,-0.200000)(0.240000,-0.200000)
\psline[linewidth=\AxesLineWidth,linecolor=GridColor](0.000000,-0.180000)(0.240000,-0.180000)
\psline[linewidth=\AxesLineWidth,linecolor=GridColor](0.000000,-0.160000)(0.240000,-0.160000)

{ \footnotesize 
\rput[t](0.000000,-0.342739){$0$}
\rput[t](2.000000,-0.342739){$2$}
\rput[t](4.000000,-0.342739){$4$}
\rput[t](6.000000,-0.342739){$6$}
\rput[t](8.000000,-0.342739){$8$}
\rput[t](10.000000,-0.342739){$10$}
\rput[t](12.000000,-0.342739){$12$}
\rput[t](14.000000,-0.342739){$14$}
\rput[t](16.000000,-0.342739){$16$}
\rput[t](18.000000,-0.342739){$18$}
\rput[t](20.000000,-0.342739){$20$}
\rput[r](-0.240000,-0.340000){$-0.34$}
\rput[r](-0.240000,-0.320000){$-0.32$}
\rput[r](-0.240000,-0.300000){$-0.3$}
\rput[r](-0.240000,-0.280000){$-0.28$}
\rput[r](-0.240000,-0.260000){$-0.26$}
\rput[r](-0.240000,-0.240000){$-0.24$}
\rput[r](-0.240000,-0.220000){$-0.22$}
\rput[r](-0.240000,-0.200000){$-0.2$}
\rput[r](-0.240000,-0.180000){$-0.18$}
\rput[r](-0.240000,-0.160000){$-0.16$}
} 

\psframe[linewidth=\AxesLineWidth,dimen=middle](0.000000,-0.340000)(20.000000,-0.160000)

{ \small 
\rput[b](10.000000,-0.360000){
\begin{tabular}{c}
Number of DMs\\
\end{tabular}
}

\rput[t]{90}(-3,-0.250000){
\begin{tabular}{c}
$\log_{10}$ $P_\mathrm{E}$\\
\end{tabular}
}
} 

\newrgbcolor{color292.0023}{0  0  0}
\psline[plotstyle=line,linejoin=1,showpoints=true,dotstyle=B|,dotsize=\MarkerSize,linestyle=none,linewidth=\LineWidth,linecolor=color292.0023]
(1.000000,-0.163019)(2.000000,-0.174964)(3.000000,-0.182501)(4.000000,-0.187956)(5.000000,-0.192194)(6.000000,-0.195588)
(7.000000,-0.198459)(8.000000,-0.200935)(9.000000,-0.203079)(10.000000,-0.204955)(11.000000,-0.206629)
(12.000000,-0.208099)(13.000000,-0.209504)(14.000000,-0.210701)(15.000000,-0.211902)(16.000000,-0.212965)
(17.000000,-0.213959)(18.000000,-0.214884)(19.000000,-0.215739)(20.000000,-0.216525)

\newrgbcolor{color293.0018}{0  0  0}
\psline[plotstyle=line,linejoin=1,showpoints=true,dotstyle=B+,dotsize=\MarkerSize,linestyle=none,linewidth=\LineWidth,linecolor=color293.0018]
(1.000000,-0.163019)(2.000000,-0.178421)(3.000000,-0.190036)(4.000000,-0.199420)(5.000000,-0.207398)(6.000000,-0.214314)
(7.000000,-0.220476)(8.000000,-0.225921)(9.000000,-0.230697)(10.000000,-0.235226)(11.000000,-0.239276)
(12.000000,-0.243060)(13.000000,-0.246340)(14.000000,-0.249414)(15.000000,-0.252200)(16.000000,-0.254769)
(17.000000,-0.257196)(18.000000,-0.259479)(19.000000,-0.262410)(20.000000,-0.264641)

\newrgbcolor{color294.0018}{0  0  0}
\psline[plotstyle=line,linejoin=1,showpoints=true,dotstyle=Bsquare,dotsize=\MarkerSize,linestyle=none,linewidth=\LineWidth,linecolor=color294.0018]
(1.000000,-0.163019)(2.000000,-0.179718)(3.000000,-0.192906)(4.000000,-0.204279)(5.000000,-0.214494)
(6.000000,-0.223887)(7.000000,-0.232659)(8.000000,-0.240940)(9.000000,-0.248819)(10.000000,-0.256363)
(11.000000,-0.263621)(12.000000,-0.270632)(13.000000,-0.277427)(14.000000,-0.284029)(15.000000,-0.290460)
(16.000000,-0.296737)(17.000000,-0.302873)(18.000000,-0.308883)(19.000000,-0.314775)(20.000000,-0.320559)

\newrgbcolor{color295.0018}{0  0  0}
\psline[plotstyle=line,linejoin=1,showpoints=true,dotstyle=Basterisk,dotsize=\MarkerSize,linestyle=none,linewidth=\LineWidth,linecolor=color295.0018]
(1.000000,-0.163019)(2.000000,-0.179208)(3.000000,-0.192059)(4.000000,-0.202940)(5.000000,-0.212469)(6.000000,-0.221053)
(7.000000,-0.229001)(8.000000,-0.236347)(9.000000,-0.243212)(10.000000,-0.249646)(11.000000,-0.255707)
(12.000000,-0.261457)(13.000000,-0.266883)(14.000000,-0.272134)(15.000000,-0.277037)(16.000000,-0.281747)
(17.000000,-0.286258)(18.000000,-0.290645)(19.000000,-0.294735)(20.000000,-0.298691)

\newrgbcolor{color296.0018}{0  0  0}
\psline[plotstyle=line,linejoin=1,showpoints=true,dotstyle=Bpentagon,dotsize=\MarkerSize,linestyle=none,linewidth=\LineWidth,linecolor=color296.0018]
(1.000000,-0.163019)(2.000000,-0.179273)(3.000000,-0.192397)(4.000000,-0.203842)(5.000000,-0.213817)(6.000000,-0.223008)
(7.000000,-0.231436)(8.000000,-0.239427)(9.000000,-0.246953)(10.000000,-0.254145)(11.000000,-0.260982)
(12.000000,-0.267606)(13.000000,-0.273844)(14.000000,-0.280006)(15.000000,-0.285754)(16.000000,-0.291579)
(17.000000,-0.297053)(18.000000,-0.302596)(19.000000,-0.307682)(20.000000,-0.312828)

{ \scriptsize  
\rput(6,-0.312){%
\psshadowbox[framesep=0pt,shadowsize=0pt, linewidth=\AxesLineWidth]{\psframebox*{\begin{tabular}{l}
\Rnode{a1}{\hspace*{0.0ex}} \hspace*{0.7cm} \Rnode{a2}{~~Proposed method rate-one} \\
\Rnode{a3}{\hspace*{0.0ex}} \hspace*{0.7cm} \Rnode{a4}{~~Proposed method rate-two} \\
\Rnode{a7}{\hspace*{0.0ex}} \hspace*{0.7cm} \Rnode{a8}{~~Proposed method rate-three} \\
\Rnode{a9}{\hspace*{0.0ex}} \hspace*{0.7cm} \Rnode{a10}{~~Proposed method rate-four} \\
\Rnode{a5}{\hspace*{0.0ex}} \hspace*{0.7cm} \Rnode{a6}{~~Optimum linear detector} \\
\end{tabular}}
\ncline[linestyle=none,linewidth=\LineWidth,linecolor=color292.0023]{a1}{a2} \ncput{\psdot[dotstyle=B|,dotsize=\MarkerSize,linecolor=color292.0023]}
\ncline[linestyle=none,linewidth=\LineWidth,linecolor=color293.0018]{a3}{a4} \ncput{\psdot[dotstyle=B+,dotsize=\MarkerSize,linecolor=color293.0018]}
\ncline[linestyle=none,linewidth=\LineWidth,linecolor=color294.0018]{a5}{a6} \ncput{\psdot[dotstyle=Bsquare,dotsize=\MarkerSize,linecolor=color294.0018]}
\ncline[linestyle=none,linewidth=\LineWidth,linecolor=color295.0018]{a7}{a8} \ncput{\psdot[dotstyle=Basterisk,dotsize=\MarkerSize,linecolor=color295.0018]}
\ncline[linestyle=none,linewidth=\LineWidth,linecolor=color296.0018]{a9}{a10} \ncput{\psdot[dotstyle=Bpentagon,dotsize=\MarkerSize,linecolor=color296.0018]}
}%
}%
} 

{ \small 
\newrgbcolor{color208.0018}{0  0  0}
\uput{0pt}[0](14.067791,-0.197396){%
\psframebox[framesep=1pt,fillstyle=solid,linestyle=none,linewidth=0.5pt]{\begin{tabular}{@{}c@{}}
$ \mathcal{E}_q= -10$ dB\\[-0.3ex]
\end{tabular}}}
} 

\end{pspicture}%
%
\caption{Comparison of error probability performance of tandem networks with different channel rates and for different number of DMs after three iterations of design with unconstrained tandem network, for quaternary hypothesis testing problem and $\mathcal{E}_q=-10$ dB.}
\label{fig:M4}
\end{figure}
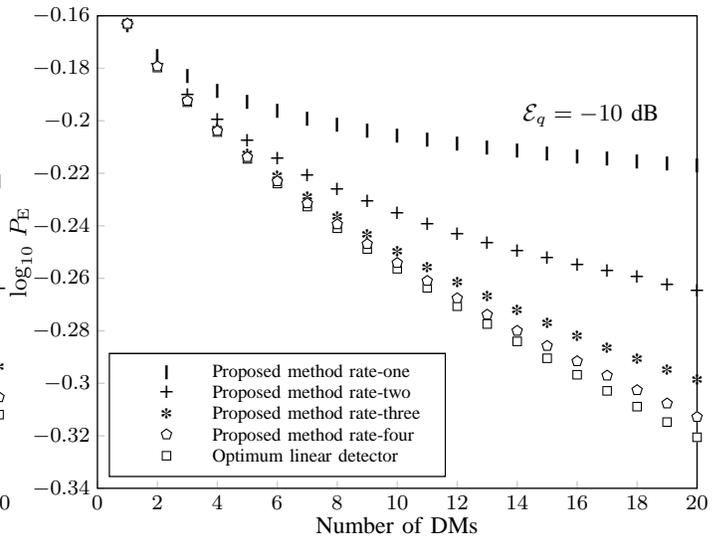

Considering the same setup as for the binary hypothesis test, we found the error probability performance of the designed tandem networks using the proposed algorithm with different channel rates for the ternary and quaternary hypothesis testing problems. We also include the performance of the unconstrained linear detector which is optimum for this problem. It is straightforward to show that this detector is just a multi-level threshold test applied to the sum in \eqref{eq:optimumlinear}, and that the error probability of the detector is equal to
\[P_\mathrm{E}=\frac{2(M-1)}{M}\,\mathcal{Q}\!\left(\frac{a\sqrt{N}}{\sigma\left(M-1\right)} \right)\,.\]
The methods of Swaszek \cite{Swa93} and Cover's \cite{Cov69} do not straightforwardly extend to the case of $M$-ary hypothesis testing. In Fig.~\ref{fig:M3} and Fig.~\ref{fig:M4}, the error probability performance of the designed tandem networks are shown for different channel rates and different number of DMs. Increasing the channel rates can significantly increase the performance of the network for $M$-ary hypothesis testing problems as in the binary case.

\section{Conclusion}\label{sec:conc}
In the context of decentralized hypothesis testing in tandem networks, we have proposed an iterative numerical algorithm which cyclically improves the performance of the network in terms of the error probability. Introducing a \emph{restricted model}, we have shown that it is possible to update the decision function of each node together with the fusion function, while all the other peripheral nodes in the network are modeled as a Markov chain.

In this paper, we have considered the hypothesis testing problem in tandem networks which is of interest since it provides a tool for the study of other more complicated topologies, like tree topologies. It can also be relevant in topologies where a single node with $m$ bit memory makes observation at different time periods and at each time makes a decision based on its current observation and a previous decision which is stored in memory, and updates its memory with the new decision.

In our model, we have assumed multi-bit communication between the sensors for the general $M$-ary hypothesis testing problem while the observations at the sensors are, conditioned on the true hypothesis, independent. The $M$-ary hypothesis test in tandem networks when the sensors make an $M$-ary decision and have conditionally dependent observations, was recently studied in \cite{PenV12}. However the problem of multi-bit communication (not necessarily $M$-ary) and conditionally dependent observations still remains open, and an extension of our work could include the design of tandem networks for conditionally dependent observations.

\bibliographystyle{IEEEtran}
\bibliography{Ref}

\end{document}